%% file: main.tex
\documentclass[amsmath, amssymb, preprint]{revtex4-2}
\usepackage[utf8]{inputenc}
\usepackage{graphicx} 

\usepackage[hidelinks]{hyperref}
\usepackage{xcolor}
\date{\today}
\include{Shortcuts}

\begin{document}
\preprint{ }

\title{Fueling limits in a cylindrical viscosity-limited reactor}
\include{Authors}

\begin{abstract}
    Recently, a method to achieve a ``natural hot-ion mode" was suggested, by utilizing ion viscous heating in a rotating plasma with a fixed boundary. We explore the steady-state solution to the Braginskii equations and find the parameter regime in which a significant temperature difference between ions and electrons can be sustained in a driven steady state. The threshold for this effect occurs at $\rho_i\gtrsim0.1R$. An analytic, leading order low flow solution is obtained, and a numerical, moderate Mach number $M\lesssim2$ is investigated. The limitation is found to be at moderate Mach numbers.
\end{abstract}
\maketitle

\noindent
\maketitle

\section{Introduction}~\label{sec:1}

Magnetic plasma confinement assisted by rotation has been explored in several configurations, such as mirrors \cite{Lehnert1971,Bekhtenev1980,Hassam1999} and toroidal devices\cite{Rax2017,Ochs2017ii}. Rotating mirrors in particular are receiving renewed interest, leading to new experimental devices in the near future\cite{Ellis2001,Ellis2005,Teodorescu2010}. Plasma mass filters\cite{Rax2016,Zweben2018,Ochs2017iii,Gueroult2012MCMF,Gueroult2015,Gueroult2019ii,Fetterman2011b,Bonnevier1966,Bonnevier1971} are another rotating plasma application in which density gradients are of particular importance, and are similar to rotating mirrors in many respects.

Sufficiently long mirrors may be analyzed using classical transport theory. Radial cross field ion currents~\cite{Kolmes2019,Rax2019} in such devices appear to be an attractive fueling method, as they can induce rotation in the plasma due to their interaction with the magnetic field. The hydrodynamic variables - densities, momenta and pressures - in such configurations can be asymptotically solved for\cite{Rubin2021} or numerically integrated using a variety of tools, such as the MITNS: Multiple-Ion Transport Numerical Solver code\cite{Kolmes2021}. 

Nuclear fusion in magnetic devices is realized by confinement of hot ions for sufficient time\cite{Lawson1957}. In these devices, the plasma confinement is often limited by the total plasma pressure, which sums the electron and ion pressures. As such, a hot-ion mode is preferable, as it can produce more fusion power for the same magnetic field strength\cite{Clarke1980}, in addition to a decrease in energy radiation  losses through electrons.

Plasma heating can be accomplished using a variety of methods. One proposed method is to use viscous heating due to sheared rotation. \citet{Kolmes2021ii} showed how heat dissipation channels in an axisymmetric cylindrical plasma could preferentially heat the ion population. Such a configuration may be realized by a radial flow of fuel ions into the hot center of the cylinder, and the removal of ash ions by a fast process other than classical transport - such as $\alpha$-channeling\cite{Fisch1992,Fisch1994}. The heating is the result of the viscous dissipation of the ion fluid, i.e., the rate of work done by the viscous stress times strain-rate. The electron fluid viscous stress, and the resultant heating rate, is smaller by a factor of $(m_e/m_i)^{3/2}$.

\citet{Kolmes2021ii} predicted that the ion temperature could be higher than the electron temperature, and suggested that the temperature difference could be large in cases with sufficiently large radial influxes of fuel ions and high radial electric fields. However, that paper left open the question of what kinds of radial influxes and fields could be self-consistently supported, and of precisely what are the density, velocity, and pressure profiles in this case. 
The present paper addresses that question. 
One of the key results that follows from this calculation is that there are nontrivial limitations on these hot-ion-mode solutions.


In this paper, we explore a particular solution to the proposed concept. We consider a one-dimensional Braginskii fluid model\cite{Braginskii1965} to explore the nonlinear effects due to the density and temperature dependence of the viscosity and heat conductivities for a uniform volumetric charge extraction, i.e. a radial ion current.  We compare analytic solutions using constant coefficients to the full numerical non-linear solution.

Because the Braginskii fluid model is the most usual model, it serves well our purposes here. We do note that several authors published corrections to the transport coefficients\cite{Epperlein1986,Simakov2022,Jeong-Young2013,Sadler2021,Davies2021}. In addition, a magnetic mirror machine would have a distribution function that contains voids where particles are not confined. In the magnetic mirror case, it is not clear that Braginskii, or other closures that expand the distribution function in a polynomial basis, produce the precise transport coefficients. However, the effects discussed in this work do not rely on the specific transport coefficients.


The limitation on the viscous heating arises because the radial ion current into the center of the cylinder leads to a large rotation of the plasma, and the centrifugal force caused by this rotation empties out the density at the core. The quadratic dependence of the viscosity on the density means that large angular velocity gradients are required to produce the viscous shear needed to balance the torque produced by the radial current and the magnetic field - leading to progressively larger rotation. This effect limits the amount of charge extraction possible in this configuration. The viscous heating itself, which increases the plasma temperature, further reduces the viscosity coefficient. Of course, diverging angular velocities are not physical, and are the result of an attempt to balance a finite torque when the viscosity coefficient approaches 0. In a physical system, the torque would be limited.

As a result, the proposed ``natural hot-ion mode" is limited. Beyond a certain radial ion flux, the nonlinearities in the viscosity coefficient would cut off the shear stress in the plasma. This limitation in shear limits the viscous heating; the limit depends on the magnetization of the plasma. We calculate a magnetization threshold above which the viscous heating is small compared to the rate of temperature equilibration between species.  

This paper is organized as follows: In Sec. \ref{sec:2} we present the nondimensionalized two-fluid equations. In Sec. \ref{sec:3} we present the low-flux approximate solution, assuming constant coefficients. In Sec. \ref{sec:4} we discuss the deviation of the full nonlinear (variable coefficients) solution from the linear approximation. 

\section{Model}~\label{sec:2}

Consider an axisymmetric, infinitely long plasma cylinder, in equilibrium, such that $\pdv{}{t} = \pdv{}{z} = \pdv{}{\theta} = 0$, with a constant axial magnetic field. The plasma species to be considered, for simplicity, are fuel ions and electrons. This setting might be realizable in a steady-state rotating mirror machine, which is fueled radially rather than axially, and in which fusion ash is removed quickly radially using $\alpha$-channeling, before it can interact with the fuel ions or the electrons. It is assumed the $\alpha$-channeling does not affect electrons. 

In steady state, fuel ions fuse, the ash is removed, and more fuel is supplied continuously from the outer edge, while the electrons have no average radial velocity. The ion sink term produces in steady state an inward-flowing ion current. The $\br \times (\bj\times\bB)$ torque due to this current induces rotation in the plasma. The plasma rotation, in addition to the other radial forces acting on it, determines the steady-state density profile. 

The radial expulsion of the fusion ash produces the opposite $\br \times (\bj\times\bB)$ torque on the ash ions. However, the fusion ash is kept at such a low density, using $\alpha$-channeling, that its collisional interactions with other plasma constituents can be ordered out of the momentum and energy equations. A density ratio of  $n_{a} / n_{i} \sim \rho_*^2$ is sufficient, with the small parameter $\rho_*$ being the normalized ion Larmor radius, also defined later.

In this work, we consider only classical transport effects in order to determine the temperatures. In many real-life plasmas, other effects contribute to the energy balance - examples include radiative cooling, and RF heating. These, and energy exchange with fusion ash may even be the dominant mechanisms, over and above classical transport effects. However, one purpose of this paper is to determine the merit of the proposed natural ion mode, which cannot be separated from the other effects considered here

For each fluid, the continuity, radial momentum, angular momentum, and pressure equations are,
\begin{gather}
    \pdv{n_s}{t} + \frac{1}{r}\pdv{}{r}r n_s v_{rs} = s_s,~\label{eq:continuity}
\end{gather}
\begin{multline} 
    \pdv{}{t}(m_sn_sv_{rs}) + \frac{1}{r}\pdv{}{r}rm_sn_sv_{rs}^2 +\pdv{p_s}{r}+(\nabla \cdot \pi_s)_r= \\s_s m_s v_{sr}^{src}+m_sn_sr\omega_s^2+Z_sen_s\left(E_r+r\omega_s B_z\right)+\sum_{s'}R_{ss'r},~\label{eq:radial_momentum}
\end{multline}
\begin{multline} 
    \pdv{}{t}(m_sn_sr^2\omega_s) + \frac{1}{r}\pdv{}{r} (r^3m_sn_s\omega_sv_{rs})   =r s_s m_s v_{s\theta}^{src}\\-Z_s e r n_s v_{rs} B_z-r(\nabla \cdot \pi_s)_\theta+r\sum_{s'}\left(R_{ss'\theta}+f_{ss'\theta}\right),~\label{eq:l}
\end{multline}
\begin{multline}
    \frac{3}{2}\pdv{p_s}{t} + \frac{1}{r}\pdv{}{r}r\left( q_{r s} + \frac{5}{2} p_s v_{r s}\right) + \pi_s : \nabla \bv_s\\ = v_{rs}  \pdv{p_s}{r}  + \sum_{s'}  3\frac{ m_s n_s \nu_{ss'}}{m_s + m_{s'}} (T_{s'} - T_s)+\frac{3}{2}s_sT^{src}_s\\+\frac{1}{2}m_s s_s\left[\left(v^{src}_{rs}-v_{rs}\right)^2+\left(v^{src}_{\theta s}-v_{\theta s}\right)^2\right] \\+ \sum_{s'} \Bigg[ \frac{m_{s'}}{m_s + m_{s'}}(\bv_{s'} - \bv_s) \cdot \left( \bR_{ss'}+\mathbf{f}_{ss'}\right)\Bigg].~\label{eq:pressure}
\end{multline}
Quantities with a subscript $s$ represent a species-dependent quantity, with the index $s$ representing ions, $s=i$, or electrons, $s=e$. 

Number density, velocity, pressure, and temperature are denoted by $n$, $v$, $p$ and $T$ respectively, with $p_s=n_sT_s$. The quantity $s_s$ is a particle source or sink for species $s$. Time and radius (spatial coordinate) are denoted by $t$ and $r$. Particle mass and charge number are denoted by $m$ and $Z$, while the elementary charge is denoted by $e$. The (radial) electric field and (axial) magnetic field are denoted by $E$ and $B$. The constant magnetic field assumption can be construed to stem from a low plasma $\beta\doteq\frac{2\mu_0 p}{B^2}$, with $\mu_0$ being the permeability of vacuum. The value used for the numerical simulations of the nonlinear equations is  $\beta = 0.002$.

If the particle source is negative, as a sink term, the source temperature $T_s^{src} = T_s$ and velocity $\bv_s^{src} = \bv_s$. If the source term is positive, $T_s^{src}$ and $\bv_s^{src}$ need to be specified.


The friction body force $\bR_{ss'}$ and the thermal friction (``Nernst") body force $\mathbf{f}_{ss'}$ between species $s$ and $s'$ are expressed as
\begin{align}
    \bR_{ss'} &= m_s n_s \nu_{ss'}(\bv_{s'}-\bv_{s}),\\
    \mathbf{f}_{ss'} &= \frac{3}{2}\frac{m_s n_s \nu_{ss'}}{Z_s Z_{s'} e B}\hat b\times\frac{Z_{s'}m_{s'}T_s\nabla T_s - Z_{s}m_{s}T_{s'}\nabla T_{s'} }{m_s T_{s'}+m_{s'}T_s}\label{eq:Nernst},
\end{align}
where $\hat b$ is a unit vector in the direction of the magnetic field. In the case of an electron-ion plasma, the thermal friction can be approximated as
\begin{gather}
    \mathbf{f}_{ie} =  \frac{3}{2}\frac{Z_i n_i \nu_{ie}}{\Omega_i}\pdv{T_e}{r}\hat \theta.
\end{gather}
The resultant viscous stress tensor divergence, using the \citet{Braginskii1965} closure, is \cite{Kolmes2021ii}
\begin{align}
    \nabla \cdot \pi_s =& \pdv{}{r}\left[\frac{\eta_{0s}}{3r}\pdv{}{r}\left(rv_{rs}\right)\right]\hat r\nonumber\\-\frac{1}{r^2}&\pdv{}{r}r^3\left[\eta_{1s}  \pdv{}{r}\left( \frac{v_{rs}}{r} \right)+\eta_{3s} \pdv{\omega_s}{r} \right] \hat r\nonumber\\ -\frac{1}{r^2}&\pdv{}{r}  r^3 \left[\eta_{1s}  \pdv{\omega_s}{r}-\eta_{3s}  \pdv{}{r}\left( \frac{v_{rs}}{r} \right) \right]\hat \theta\label{eq:divpi}.
\end{align}

The viscosity coefficients for an electron-ion plasma, with $Z=1$ are
\begin{gather}
    \eta_{0i} = 0.96\sqrt{2}\frac{p_i}{\nu_{ii}},\\
    \eta_{1i} =\frac{3}{10\sqrt{2}} \frac{p_i \nu_{ii}}{\Omega_i^2},\\
    \eta_{3i} = \frac{p_i}{2\Omega_i},\\
    \eta_{0e} = 0.73\frac{p_e}{\nu_{ei}},\\
    \eta_{1e} = 0.51\frac{p_e \nu_{ei}}{\Omega_e^2},\\
    \eta_{3e} = \frac{p_e}{2\Omega_e},
\end{gather}
where $\Omega_s$ denotes the signed Larmor frequency for species $s$. 
The viscous heating term is expressed as
\begin{multline}
    \pi_s : \nabla \bv_s=-\frac{\eta_{0s}}{3} \left(\frac{1}{r}\pdv{}{r}\left(r v_{rs} \right)\right)^2\\-\eta_{1s} \left(\left(r \pdv{}{r}\left(\frac{v_{rs}}{r}\right)\right)^2+\left(r \pdv{\omega_s}{r}\right)^2 \right).
\end{multline}
The heat conduction $q_r$ is
\begin{gather}
    q_{rs} = -\kappa_{s}\pdv{T_s}{r} - \kappa_{ss'}^{\omega}r\left(\omega_{s'}-\omega_{s}\right), 
\end{gather}
with heat diffusivities,
\begin{gather}
    \kappa_{i} =\sqrt{2}\frac{ p_i\nu_{ii}}{m_i\Omega_i^2},
    \\
    \kappa_{e} = 4.66\frac{ p_e\nu_{ei}}{m_e\Omega_e^2},
    \\
    \kappa_{ie}^{\omega} \approx 0,\\
    \kappa_{ei}^{\omega} = \frac{3}{2}\frac{p_e\nu_{ei}}{\Omega_e}.
\end{gather}
The Coulomb collision frequency between particles of species $s$ and $s'$ is expressed as\cite{Fundamenski2007}
\begin{multline}
    \nu_{ss'} = \frac{\sqrt{2}e^4 \log \Lambda_{ss'}}{12\pi^{3/2}\varepsilon_0^2}\cdot\\Z_s^2 Z_{s'}^2\sqrt{\frac{m_{s'}}{m_s}\frac{1}{m_s+m_{s'}}}\left(\frac{m_s+m_{s'}}{m_sT_{s'}+m_{s'}T_s}\right)^{3/2}n_{s'}.
\end{multline}
with $\log \Lambda_{ss'}$ being the Coulomb logarithm and $\varepsilon_0$ the permittivity of free space. Note that this definition of collision frequencies for like-species collisions differs from its expression in Braginskii by a factor of $1/\sqrt{2}$. The corresponding transport coefficients incorporate a $\sqrt{2}$ factor, so the transport itself is unchanged.

In steady state, the particle flux
\begin{gather}
    \Gamma_s \doteq r n_s v_{rs} = \Gamma_s(R) + \int_{R}^{r} rs_s \mathrm{d}r,
\end{gather}
is the quantity driving the system. Here we use $R$ as the outer radius of the cylinder.

\subsection{Source terms and boundary conditions}

The source terms and fluxes for the ions and electrons,
\begin{gather}
    \Gamma_e \equiv 0,\\
    s_i = \mathrm{const.}<0,\\
    \Gamma_i = \frac{s_i}{2}r^2,
\end{gather}
correspond to a constant ion sink $s_i<0$, understood to be a fusion process, and no radial motion of electrons.

We choose to use a uniform particle sink because we would like to compare a constant coefficient solution to the same case with variable coefficients. In the comparison, we want to be clear that the nonlinear behavior originates in the nonlinear transport coefficients rather than from the nonlinear source term.

The boundary conditions for the plasma are:
\begin{gather}
    \omega_i(R) = 0,\label{eq:wibc}\\
    \omega_e(R) = 0,~\label{eq:ewbc}\\
    Z_i n_i(R) =  n_e(R) =Z_i n_0,\label{eq:n0bc}\\
    T_i(R) = T_e(R)=T_0,\label{eq:T0bc}\\
    \omega_i'(0)=\omega_e'(0)=0,\label{eq:wcybc}\\
    n_i'(0)=n_e'(0)=0,\label{eq:ncybc}\\
    T_i'(0) =T_e'(0) = 0.\label{eq:tcybc}
\end{gather}
The integrated viscous heating of the plasma, to leading order\cite{Kolmes2021ii}, is
\begin{gather}
    \int_0^{R}\pi:\nabla \bv r \mathrm{d}r = \int_0^{R}v_{\theta}(\nabla\cdot\pi)_\theta r \mathrm{d}r + \left[r v_\theta \pi_{r\theta}\right]_{r=R}.\label{eq:visc_hea}
\end{gather}
The choice of the boundary conditions (\ref{eq:wibc}) and (\ref{eq:ewbc}) drops the boundary term in (\ref{eq:visc_hea}), which is the work done by the boundary on the plasma. Boundary conditions (\ref{eq:n0bc}) and (\ref{eq:T0bc}) are a choice of normalization, where the equality of the ion and electron temperatures might require further justification (see appendix). The boundary conditions (\ref{eq:wcybc}), (\ref{eq:ncybc}), and (\ref{eq:tcybc}) are the result of the cylindrical geometry.

\subsection{Nondimensionalization}
Nondimensionalizing the equations of motion allows us to factor out small parameters for use in an asymptotic expansion. 
Denoting $X = X_0 \tilde X$, with $X_0$ being a reference quantity: 
\begin{gather}
    m_0 \doteq m_p,\\
    v_0 \doteq \sqrt{T_0/m_p},\\
    r_0 \doteq R,\\
    \Gamma_0 \doteq R n_0 v_0,\\
    \nu_0 \doteq \frac{\sqrt{2}e^4 \log \Lambda}{12\pi^{3/2}\epsilon_0}\frac{n_0}{T_0^{3/2}m_0^{1/2}}\\
    \eta_{00}\doteq \frac{n_0 T_0}{\nu_0}\\
    \eta_{10} \doteq \frac{n_0T_0\nu_0}{\Omega_{p0}^2} = \epsilon^2\eta_{00},\\
    \eta_{30} \doteq \frac{n_0T_0}{\Omega_{p0}} = \epsilon\eta_{00},\\
    R_{ss'0} \doteq m_0 n_0 \nu_0 v_{0},\\
    f_{ss'0} \doteq \frac{n_0 \nu_0 T_0}{\Omega_{p0} R} = m_0 n_0 \nu_0 v_0 \rho_{*},\\
    \kappa_{s0} \doteq \frac{n_0T_0\nu_0}{m_0\Omega_{p0}^2} = \frac{\epsilon^2}{m_0}\eta_{00},\\
    \kappa_{ss'0}^\omega \doteq \frac{n_0T_0\nu_0}{\Omega_{p0}}.
\end{gather}
In this paper, reference quantities are chosen as their value at the outer radius.

This fluid closure features two small parameters, the normalized ion Larmor radius, $\rho_{*} \doteq  v_0/\Omega_{p0}R$, and the ratio of collision frequency to Larmor frequency $\epsilon \doteq \nu_0/\Omega_{p0}$, which is the inverse Hall parameter $\epsilon = 1/C_H$. We use the reference quantities to define the values of these constants, i.e., we use them as constants rather than as functions of radius. When dealing with electron-ion plasma, a third small parameter is present, the square root of the electron-to-proton mass ratio, $\sme$.
An asymptotic expansion for a parameters $\tilde X$, in powers of the small parameters $\rho_*$ and $\epsilon$ would be denoted by $\tilde X = \sum_{\alpha, \beta}\rho_{*}^{\alpha}\epsilon^\beta \tilde X^{(\alpha,\beta)}$.

The steady-state dimensionless angular momentum equation for a single fluid species is,
%
\begin{multline}
     \tm_s  \dv{}{\tr}\tilde\Gamma_s\tr^2 \tomega_s +\frac{1}{\rho_{*}}Z_s  \tilde\Gamma_s \tB_z\tr =\tr^2 \ts_s \tm_s \tomega_{s}\\ \rho_{*}\dv{}{\tr}\tr^3 \left[\epsilon\tilde \eta_{1s}\dv{\tomega_s}{\tr}-\frac{\tilde\eta_{s3}\tilde \Gamma_s}{\tr^2 \tn_s}\left(\dv{\ln(\tilde \Gamma_s)}{\tr}-\frac{2}{\tr }-\dv{\ln (\tn_s)}{\tr}\right)\right]
    \\+\epsilon\tr^2\sum_{s'}\left(\frac{1}{\rho_{*}}\tilde R_{ss'\theta} +\tilde f_{ss'\theta}\right).~\label{eq:2ndorderomegai}
\end{multline}
The steady-state dimensionless radial force-balance equation for a single fluid species is,
\begin{multline} 
	\dv{\tp_s}{\tr}=\frac{1}{\rho_{*}}Z_s\tn_s\left(\tE_r+\tr\tomega_s \tB\right)+\frac{\epsilon}{\rho_{*}}\sum_{s'}\tilde R_{ss'r}\\+\tm_s \ts_s \frac{\tilde \Gamma_s}{\tr \tn_s}+\tm_s\tn_s\tr\tomega_s^2+\frac{1}{\tr}\dv{}{\tr}\left(\frac{\tm_{s}\tilde \Gamma_{s}^2}{\tr\tn_s} \right)  \\+ \frac{\rho_{*}}{\epsilon}\dv{}{\tr}\left(\frac{\tilde\eta_{s0}}{3\tr}\dv{}{\tr}\left(\frac{\tilde\Gamma_s}{\tn_s}\right)\right)\\+\frac{\rho_{*}}{\tr^2}\dv{}{\tr}\tr^3\left(\epsilon\tilde \eta_{s1}  \dv{}{\tr}\left(\frac{\tilde\Gamma_s}{\tr^2 \tn_s}\right)+\tilde\eta_{s3}\dv{\tomega_s}{\tr} \right).\label{eq:vrssdim}
\end{multline}
The steady-state dimensionless temperature equation for a single fluid species is,
\begin{multline}
    -\frac{1}{\tr}\dv{}{\tr} \tr\left[\tilde\kappa_{s}\dv{\tT_s}{\tr} + \frac{1}{\rho_{*}}\tilde\kappa_{ss'}^{\omega}\tr\left(\tomega_{s'}-\tomega_s\right)\right] + \frac{5}{2}\frac{1}{\tr}\dv{}{\tr} \tT_s \frac{\tilde \Gamma_s}{\rho_*\epsilon} 
    \\ = \dv{\tp_s}{\tr}\frac{\tilde\Gamma_s}{\rho_{*}\epsilon \tr \tn_s}  +\frac{1}{\rho_{*}^2} \sum_{s'}  3\frac{ \tm_s \tn_s \tilde \nu_{ss'}}{\tm_s + \tm_{s'}} (\tT_{s'} - \tT_s)+\frac{3}{2}\frac{\tilde s_s\tT_s}{\rho_{*}\epsilon}
    \\ +\sum_{s'} \Bigg[ \frac{\tm_{s'}}{\tm_s + \tm_{s'}}\frac{1}{\rho_{*}}(\tilde \bv_{s'} - \tilde\bv_s) \cdot \left( \frac{1}{\rho_{*}}\tilde \bR_{ss'}+ \tilde{\mathbf{f}}_{ss'}\right)\Bigg]
    \\+\frac{1}{\epsilon^2}\frac{\tilde \eta_0}{3} \left[\frac{1}{\tr}\dv{}{\tr}\frac{\tilde \Gamma_s}{\tn_s}\right]^2+\tilde \eta_1 \tr^2 \left[\left[ \dv{}{\tr}\left(\frac{\tilde \Gamma_s}{\tr^2 \tn_s}\right)\right]^2+\left[\dv{\tomega_s}{\tr}\right]^2 \right]~\label{eq:dimlesspressure}
\end{multline}

\section{Constant coefficients solution}~\label{sec:3}
In this section, the leading order solution for the angular velocities, density and temperatures are derived, assuming $\eta_1$, $\kappa$ and $\nu_{ie}$ are constants, and do not depend on the variations in density or temperature within the domain. 
This is an approximation that holds well for slow rotation ($\tv_{\theta}\ll1$), when the density and temperature are indeed nearly uniform.

\subsection{Electron angular velocity}
For $\Gamma_e\equiv 0$ the electron angular velocity equation (\ref{eq:2ndorderomegai}) reads,
\begin{gather}
    \rho_{*}\dv{}{\tr}\left(\tr^3 \tilde \eta_{1e}\dv{\tomega_e}{\tr}\right)
    =\frac{1}{\rho_{*}} \tm_i \tn_i \tilde\nu_{ie}\tr^3(\tomega_e-\tomega_i) +\tr^2\tilde f_{ie\theta}.
\end{gather}
The electron angular velocity can be solved asymptotically, to leading and first order in $\rho_{*}$,
\begin{gather}
    \tomega_e = \tomega_i-\frac{3}{2}\frac{\rho_{*}}{\tr \tB}\dv{\tT_e}{\tr}\left(1-\frac{I_1\left(\frac{\tr}{\rho_{*}}\sqrt{\frac{ \tm_e \tn_e \tilde \nu_{ei}}{\tilde \eta_{1e}}}\right)}{I_1\left(\frac{1}{\rho_{*}}\sqrt{\frac{ \tm_e \tn_e \tilde \nu_{ei}}{\tilde \eta_{1e}}}\Big|_{\tr=1}\right)}\right)~\label{eq:ewsol},
\end{gather}
where $I_1$ is the modified Bessel function of the first kind.
This solution sets the azimuthal component of the total friction force $R_{ie\theta}+f_{ie\theta}=0$, except for a boundary layer at $\tr\approx 1$.

The electron viscosity coefficient is smaller by a factor of $\tm_e^{3/2}$ relative to the ion viscosity coefficient, and the contribution of the viscosity is of $\mathcal{O}(\rho_{*}^{2}\tm_e)$ relative to the ion angular velocity.
%
%
%
%
The ion angular velocity equation can be exactly solved if the sink term $\tr^2 \ts_s \tm_s \tomega_{s}$ is dropped. This term would turn out to be of $\mathcal{O}(\rho_*^2\epsilon)$ later. The exact solution to $\tomega_i$ (as a function of $\tomega_e$) without the sink term, with boundary conditions $ \tomega_i(1)=0$ is 
\begin{multline}
    \tomega_i = e^{\int_{1}^{\tr}\frac{\tm_i \tilde\Gamma_i}{\rho_{*}\epsilon\tilde \eta_{1i}}\frac{d\tr'}{\tr'}} \int_{1}^{r}d\tr' e^{-\int_{1}^{\tr'}\frac{\tm_i \tilde\Gamma_i}{\rho_{*}\epsilon\tilde \eta_{1i}}\frac{d\tr''}{\tr'' }}\cdot\\
    \left[\frac{Z_i}{\rho_{*}^2\epsilon}\frac{\int_{0}^{\tr'}\tilde \Gamma_i\tB\tr''d\tr''}{\tr'^3\tilde \eta_{1i}}
    +\frac{\tilde \eta_{1e}\tomega_e'}{\tilde \eta_{1i}}\right.
    \\\left.-\frac{ \tilde\Gamma_i\tilde\eta_{3i}}{\epsilon\tilde \eta_{1i}\tr'^2 \tn_i}\left[\frac{2}{\tr'}+\dv{ \ln(\tn_i)}{\tr}\right] \right].~\label{eq:omega_isol}
\end{multline}
Notice that each term in (\ref{eq:omega_isol}) is proportional to 
\begin{gather} 
    \frac{1}{\tilde \eta_{1i}} \propto \frac{\sqrt{\tT_i}}{\tn_i^2}.
\end{gather}
If $\tn_i$ were to continuously decrease from $1$ to near $0$ when moving from $\tr =1$ inwards or if the ion temperature should diverge, $\tomega_i$ would diverge. This is due to the decrease in viscosity while the magnetic field torque remains the same.   

The leading order solution for $\tomega_i$, taking $\tilde \Gamma_i=\frac{1}{2}\ts_i\tr^2$, which is the solution to the continuity equation with a uniform steady source term $\ts_i=\mathrm{const.}$, and $\tB=\mathrm{const.}$, in addition to $\tilde \eta_{1i}=\mathrm{const.}$ is,
\begin{multline}
    \tomega_i^{(0,0)} = \frac{Z_i\tB\ts_i}{8\rho_{*}^2\epsilon\tilde \eta_{1i}}e^{\frac{\ts_i \tm_i}{4\rho_{*}\epsilon\tilde \eta_{1i}}(\tr^2-1)}\cdot \int_{1}^{r} e^{-\frac{\ts_i \tm_i}{4\rho_{*}\epsilon\tilde \eta_{1i}}(\tr'^2-1)}
    \tr'
    \mathrm{d}\tr'~\label{eq:omega_isol_toy}\\
     = \frac{\tOmega_i}{4\rho_{*}}\left(e^{F_i(\tr^2-1)} -1\right),
\end{multline}
It is useful to define the variables $F_i = \frac{\ts_i \tm_i}{4\rho_{*}\epsilon\tilde \eta_{1i}}$, which is the mass flux over viscosity, and $\tOmega_i\doteq \frac{Z_i\tB}{\tm_i}$ which is the dimensionless gyro-frequency. The solution is exponentially dependent on the strength of the source term $\ts_i$. 

The azimuthal velocity in this case is
\begin{gather}
    \tv_{\theta i}^{(0,0)}= 
    \frac{\tOmega_i}{4\rho_{*}}\tr\left(e^{F_i(\tr^2-1)} -1\right).
\end{gather}
The angular velocity becomes $\mathcal{O}(1)$ if $F_i\sim\mathcal{O}(\rho_{*})$. An azimuthal Mach 1 ($\tv_{\theta i}=1$) is obtained when $F_i\approx -6\sqrt{3}\rho_{*}/\tOmega_i$, corresponding to a source term $\ts_i\sim\mathcal{O}(\rho_{*}^2\epsilon)$.
The next corrections are of $\mathcal{O}(\tm_e^{3/2})$ due to the electron viscosity and $\mathcal{O}(\rho_{*}^{2})$ due to the $\tilde \eta_{3i}$ rotation term.

\subsection{Temperatures}
The ion temperature equation, when substituting $\tilde\Gamma_i =\rho_{*}^2\epsilon \tilde \Gamma_i^{(2,1)}$, and remembering $\tomega_i\sim\mathcal{O}(1)$
\begin{multline}
    -\frac{1}{\tr}\dv{}{\tr}\left( \tr\tilde\kappa_{i}\dv{\tT_i}{\tr}\right) +\rho_{*} \frac{5}{2}\frac{1}{\tr}\dv{}{\tr} \tT_i \tilde \Gamma_i^{(2,1)} 
    \\ =3\frac{1}{\rho_{*}^2} \tn_i \tilde \nu_{ie}(\tT_e - \tT_i)+\tilde \eta_{1i} \left(\tr \dv{\tomega_i}{\tr}\right)^2 \\+\rho_{*} \dv{\tp_i}{\tr}\frac{\tilde\Gamma_i^{(2,1)}}{\tr \tn_i}  + \frac{3}{2}\rho_{*} \ts_i^{(2,1)}\tT_i
    \\ +\rho_{*}^4\frac{\tilde \eta_{0i}}{3} \left(\frac{1}{\tr}\dv{}{\tr}\frac{\tilde \Gamma_i^{(2,1)}}{\tn_i}\right)^2+\tilde \eta_{1i} \left(\rho_{*}^2\epsilon\tr \dv{}{\tr}\left(\frac{\tilde \Gamma_i^{(2,1)}}{\tr^2 \tn_i}\right)\right)^2,~\label{eq:Ti}
\end{multline}
and the electron equation,
\begin{multline}
    -\frac{1}{\tr}\dv{}{\tr} \left(\tr\tilde\kappa_{e_{eff}}\dv{\tT_e}{\tr}\right)
    = -3\frac{1}{\rho_{*}^2} \tn_i \tilde \nu_{ie}(\tT_e - \tT_i)
    \\ +\rho_{*}^2\epsilon^2 \frac{\tm_i\tilde \nu_{ie}}{\tr^2\tn_i}\tilde \Gamma_i^{(2,1)^2}+\tilde \eta_{1e} \left(\tr \dv{\tomega_e}{\tr}\right)^2
\end{multline}
with
\begin{gather}
    \tilde \kappa_{e_{eff}} = \left[4.66+\frac{9}{4}\left[1-\frac{I_1\left(\frac{\tr}{\rho_{*}}\sqrt{\frac{ \tm_e \tn_e \tilde \nu_{ei}}{\tilde \eta_{1e}}}\right)}{I_1\left(\frac{1}{\rho_{*}}\sqrt{\frac{ \tm_e \tn_e \tilde \nu_{ei}}{\tilde \eta_{1e}}}\Big|_{\tr=1}\right)}\right]\right]\frac{\tp_e\tilde \nu_{ei}}{\tm_e\tOmega_e^2}~\label{eq:kappaeeff},
\end{gather}
being the effective heat transfer coefficient for the electrons, due to the contribution of the Ettingshausen, $\tilde \kappa^\omega$ term. 
The plasma heat transfer coefficient, for $Z_i=1$, $\tm_i=1$, is $\tilde \kappa_{tot}=\tilde \kappa_i+\tilde \kappa_{e_{eff}}\approx1.16\tilde \kappa_i$, outside of the boundary layer.

This is a linear ODE for $\tT_i$, with a single term - the collisional equilibration between ions and electrons $\sim\rho_{*}^{-2}$. Thus, in the low-flow case, $F_i\sim\mathcal{O}(\rho_*)$, $\Gamma_i\sim\mathcal{O}(\rho_{*}^2\epsilon)$, the temperature difference between the ion and electron fluids is of $\mathcal{O}(\rho_{*}^2\tomega_i^2/\sme) \sim \mathcal{O}(\rho_{*}^2/\sme) $. 

To leading order, with constant coefficients, the sum of the two equations is,

\begin{gather}
    -\frac{1}{\tr}\dv{}{\tr}\left( \tr\tilde\kappa_{tot}\dv{\tT_i^{(0,0)}}{\tr}\right)
    =\tilde \eta_{1i}\left(\tr \dv{\tomega_i}{\tr}\right)^2,
\end{gather}
yielding
\begin{multline}
    \tT_i^{(0,0)} = -\frac{\tilde \eta_{1i}}{\tilde\kappa_{tot}}\frac{\tOmega_i^2e^{-2F_i}}{32\rho_{*}^2 F_i}\Bigg(\frac{1}{2}(Ei(2F_i\tr^2)-Ei(2F_i))-\ln(\tr)\\+\frac{1}{4}(e^{2F_i\tr^2}(2F_i\tr^2-3)-e^{2F_i}(2F_i-3))\Bigg)+\tT_i(1)
\end{multline}
with $\tilde \eta_{1i}/\tilde\kappa_{i} = 0.15\tm_i$ and $\tilde \eta_{1i}/\tilde\kappa_{tot} \approx 0.13\tm_i$. These ratios are related to the Prandtl number as $\tilde \eta/\tilde \kappa =\frac{2}{3}\tm \mathrm{Pr}$. The function $Ei(x)=\int_{-\infty}^{x}\frac{e^t}{t}dt$ is the exponential integral. For small values of $F_i$, such as $F_i \approx -\rho_*$, the boundary term $\tT_i(1)=1$ is the dominant term in the leading-order temperature solution.




The temperature difference between electrons and ions $\Delta T_{ei} = \tT_e-\tT_i$, is determined, to leading order, by
\begin{multline}
    -\frac{1}{\tr}\dv{}{\tr}\left( \tr\dv{\Delta T_{ei}}{\tr}\right)= -\frac{3\tn_i \tilde \nu_{ie}}{\rho_{*}^2}\left(\frac{1}{\tilde\kappa_{e_{eff}}}+\frac{1}{\tilde\kappa_{i}}\right)\Delta T_{ei}
    \\+\frac{2}{3}\left(\tm_e \mathrm{Pr}_e-\tm_i \mathrm{Pr}_i \right)\left(\tr \dv{\tomega_i}{\tr}\right)^2.~\label{eq:DeltaT}
\end{multline}
The leading order solution to this equation is 
\begin{multline}
    \Delta T_{ei} =\left.-\frac{2\rho_{*}^2}{9\tn_i \tilde \nu_{ie}} \frac{\tm_i \mathrm{Pr}_i-\tm_e \mathrm{Pr}_e}{\frac{1}{\tilde\kappa_{i}}+\frac{1}{\tilde\kappa_{e_{eff}}}}\right(\tr^2 \left[\dv{\tomega_i}{\tr}\right]^2
    \\\left.-\left[\dv{\tomega_i}{\tr}\right]^2\Bigg|_{\tr=1}\frac{I_0\left(\frac{\tr}{\rho_*}\sqrt{3\tn_i \tilde \nu_{ie}\left(\tilde\kappa_{e_{eff}}^{-1}+\tilde\kappa_{i}^{-1}\right)}\right)}{I_0\left(\frac{1}{\rho_*}\sqrt{3\tn_i \tilde \nu_{ie}\left(\tilde\kappa_{e_{eff}}^{-1}+\tilde\kappa_{i}^{-1}\right)\Big|_{\tr=1}}\right)}\right)~\label{eq:DelTeisol}
\end{multline}
with $I_0$ being the modified Bessel function of the first kind. 

Since $\tm_e \mathrm{Pr}_e<\tm_i \mathrm{Pr}_i$, the solution is negative, indeed yielding a hot-ion mode, but it also contains a pre-factor of $\rho_*^2/\sme$. 
In order to achieve a temperature difference of $\mathcal{O}(1)$, we identify a course of action, that does not suffer from the fueling limit discussed later. In any case, one has to push $\tomega$ to be as large as possible. However, the $\rho_*^2/\sqrt{\tm_e}$ pre-factor has to be dealt with as well: Use low magnetization, such that $\rho_*\gtrsim \sqrt[4]{\tm_e} \sim 0.1$. This would bring the pre-factor to be $\mathcal{O}(1)$. In this case, $\ts_i^{(2,1)}$ can remain small such that the angular velocity is still linear in $\ts_i$. The source term magnitude must then be larger than the square root of thecombined value of the reminder of (\ref{eq:DelTeisol}), which is of $\mathcal{O}(0.001)$, when substituting $\tomega_i$ as a function of $\ts_i$. In case of Braginskii coefficients for $Z_i=\tm_i=1$, this source term is $\ts_i \approx -20\rho_*^2\epsilon$. It is perhaps easier to increase $\rho_*$ rather than $\ts_i$, as the maximal temperature difference increases as $\rho_*^3 \left(\ts_i^{(2,1)}\right)^2$ at large $\rho_*$, see Figure \ref{fig:Delta_T}. In the figure, the temperature difference is $\sim0.01$ because the source term magnitude kept relatively small for numerical reasons. This Larmor radius is becoming somewhat large for a small parameter, required for the fluid approximation. Some magnetic traps operate with this Larmor radius / machine size scale.


\subsection{Density}

The density profile is determined from the radial force balance. Summing electrons and ions, taking $\tn_e = Z_i\tn_i$, and neglecting electron viscosity, the radial force balance of the plasma as a whole is
\begin{multline} 
	\dv{\tp_i}{\tr}+\dv{\tp_e}{\tr}=\frac{3}{2}Z_i\tn_i\dv{\tT_e}{\tr} +\tn_i\tr\left(\tm_i\tomega_i^2+ Z_i \tm_e\tomega_e^2\right)\\+\frac{\tm_i \tilde \Gamma_i^2}{\tr^2 \tn_i}\left(\frac{1}{\tr}+\dv{\ln(\tn_i)}{\tr}-\dv{ \ln(\tilde\Gamma_i^2)}{\tr}\right)\\+ \frac{\rho_{*}}{\epsilon}\dv{}{\tr}\left(\frac{\tilde\eta_{0i}\tilde\Gamma_i}{3\tr\tn_i}\left[\dv{\ln(\tilde\Gamma_i)}{\tr}-\dv{\ln(\tn_i)}{\tr}\right]\right)\\+\frac{\rho_{*}}{\tr^2}\dv{}{\tr}\tr^3\left(\tilde\eta_{3i}\dv{\tomega_i}{\tr}+\tm_e\tilde\eta_{3e}\dv{\tomega_e}{\tr}\right.\\\left.+\epsilon\frac{\tilde \eta_{1i}\tilde\Gamma_i}{\tr^2\tn_i} \left[\dv{\ln(\tilde\Gamma_i)}{\tr}-\dv{\ln(\tn_i)}{\tr}-\frac{2}{\tr}\right] \right).~\label{eq:pplasma}
\end{multline}
%
%
This is a second-order nonlinear ODE in $\tn_i$, and solving the leading-order terms in it would both remove the non-linearity, and reduce it to a first-order differential equation.
To leading order,
\begin{gather}
    \dv{\ln(\tn_i)}{\tr}+\frac{1-\frac{1}{2}Z_i}{1+Z_i}\dv{\ln(\tT_i)}{\tr}= 
	\frac{\tr\left(\tm_i\tomega_i^2+ Z_i \tm_e\tomega_e^2\right)}{\tT_i(1+Z_i)},
\end{gather} 
or for $Z_i=1$,
\begin{gather}
    \tn_i^{(0,0)} = \frac{1}{\sqrt[4]{\tT_i}}e^{\frac{\tm_i\tOmega_i^2}{32\rho_{*}^2}\int_{1}^{\tr}\frac{\tr'\left(e^{F_i(\tr'^2-1)}-1\right)^2}{\tT_i}d\tr'}.
\end{gather}
This density profile is hollow - matter is pushed to outer radii from the center.

\begin{figure*}
    \centering
    \includegraphics[width=\textwidth]{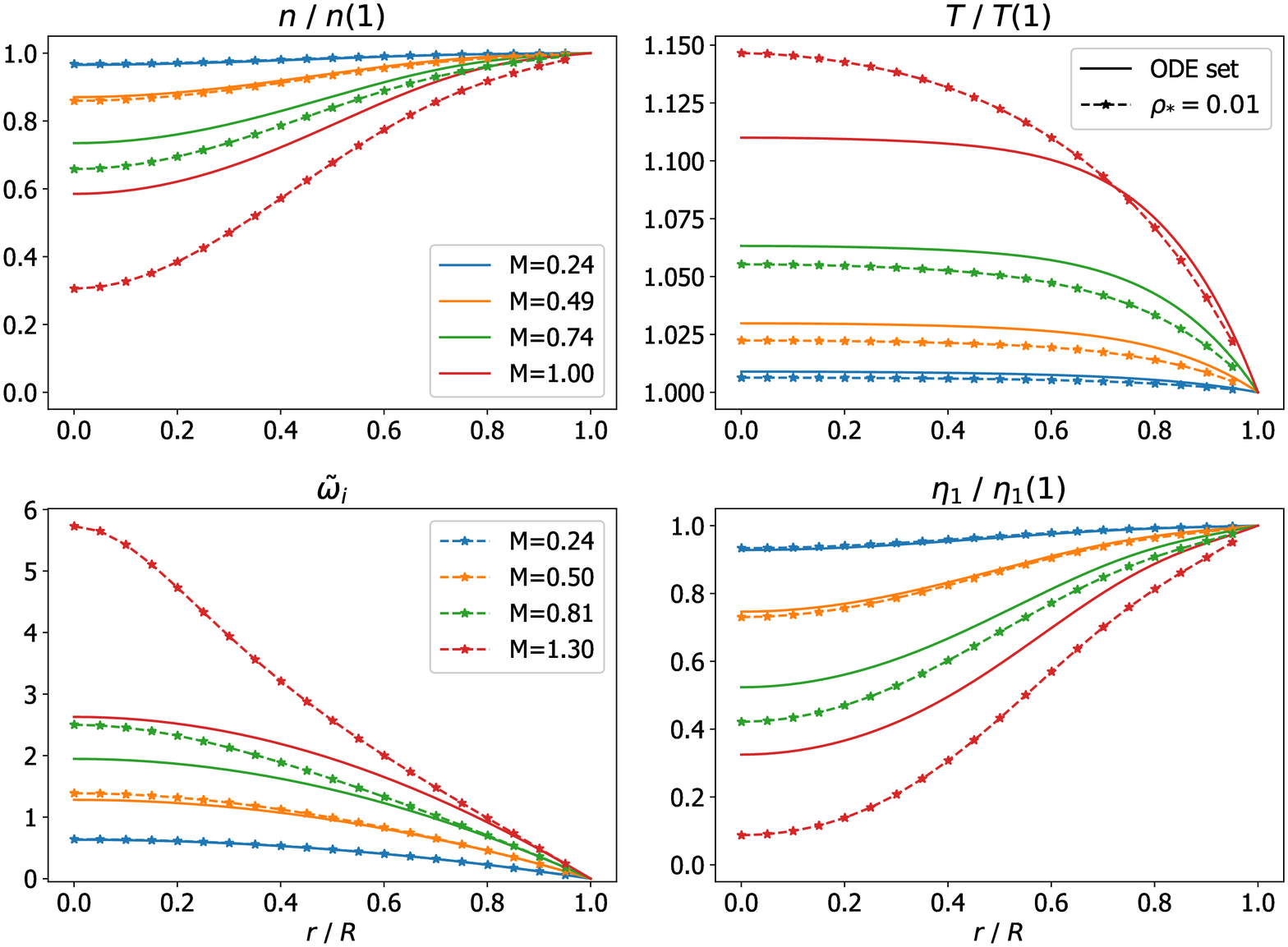}
    \caption{Comparison between solutions to the constant-coefficient leading order Braginskii equations in steady state (full line), and MITNS nonlinear results (dashed line with markers). The source term used for the ODE set and MITNS were the same. The nonlinear solution starts diverging from the leading order linear solution at Mach number of $M\approx 0.5$. At a Mach number of $M\approx1$ in the linear solution, the difference between the linear and nonlinear solutions is quite significant.}
    \label{fig:const_coeff_sol}
\end{figure*}

\subsection{Validity of the constant coefficients solution}

The viscosity profile, to leading order, is
\begin{gather}
    \tilde \eta_{1i}\propto \frac{\tn_i^2}{\sqrt{\tT_i}} = \frac{1}{\tT_i}e^{\frac{\tm_i\tOmega_i^2}{16\rho_{*}^2}\int_{1}^{\tr}\frac{\tr'\left(e^{F_i(\tr'^2-1)}-1\right)^2}{\tT_i}\mathrm{d}\tr'}.
\end{gather}
Some solutions to the constant coefficient equations are presented in Figure \ref{fig:const_coeff_sol}. Notice how $\tilde \eta_{1i} \propto \tilde \kappa_i$ drops rapidly as the source term is increased. The disagreements between the linear and nonlinear solutions are visible at Mach number of $M\approx 0.5$, and become increasingly severe.
%
%
%
%

%
\section{Variable coefficients solution}~\label{sec:4}

Braginskii's fluid model is inherently nonlinear, with collision frequencies that depend on the densities and temperatures. The diffusion coefficients (viscosities and heat transfer coefficients), present additional nonlinearity. From the linear, constant coefficients solution, it is evident that the viscosity and heat transfer coefficients drop as the magnitude of the particle flux increases. The shorthand $F_i$ would become larger at smaller radii, and the angular velocity increase beyond its constant coefficients solution values.

There are two ways in which a particle flux would fail to produce a physical solution:
\begin{enumerate}
    \item The solution breaks the ordering $\rho_i\ll R$.
    \item The solution produces a negative pressure, or is unable to satisfy both boundary conditions.
\end{enumerate}
For the first case, even a constant coefficient solution with $F_i\sim\mathcal{O}({\rho_*})$ would produce rotations that are $\tomega\sim\mathcal{O}(\rho_*^{-1})$, and temperature that is $\mathcal{O}(\rho_*^{-2})$. The Larmor radius $\rho_i \propto \sqrt{T_i}$, and increasing the temperature above its reference value by a factor of $\rho_*^{-2}$ puts us firmly in the kinetic regime, where the fluid model is inapplicable. Using a variable-coefficients solution would bring that threshold to smaller values of particle flux.



The second case is a feature of non-linear ODEs, where there is no guarantee for the existence of a boundary-value-problem solution in all cases. We can attempt solving initial value problems, and using a shooting method to pinpoint the correct values and derivatives at one boundary to hit the correct values at the other, or transform the ODEs into PDEs, and attempt to relax the solution to a steady state at finite times.
%
%
%

\begin{figure}
    \centering
    \includegraphics[width = \columnwidth]{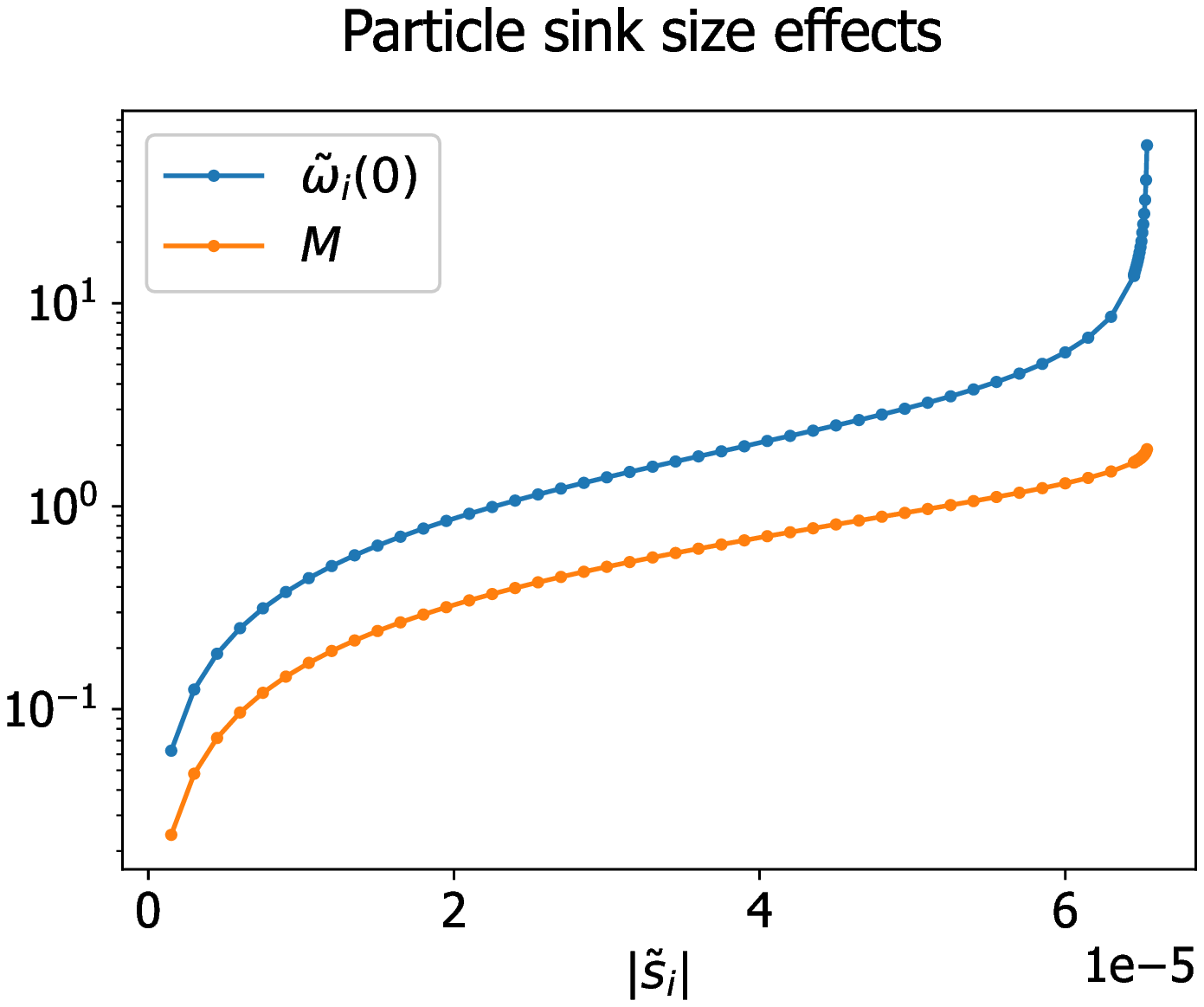}
    \includegraphics[width = \columnwidth]{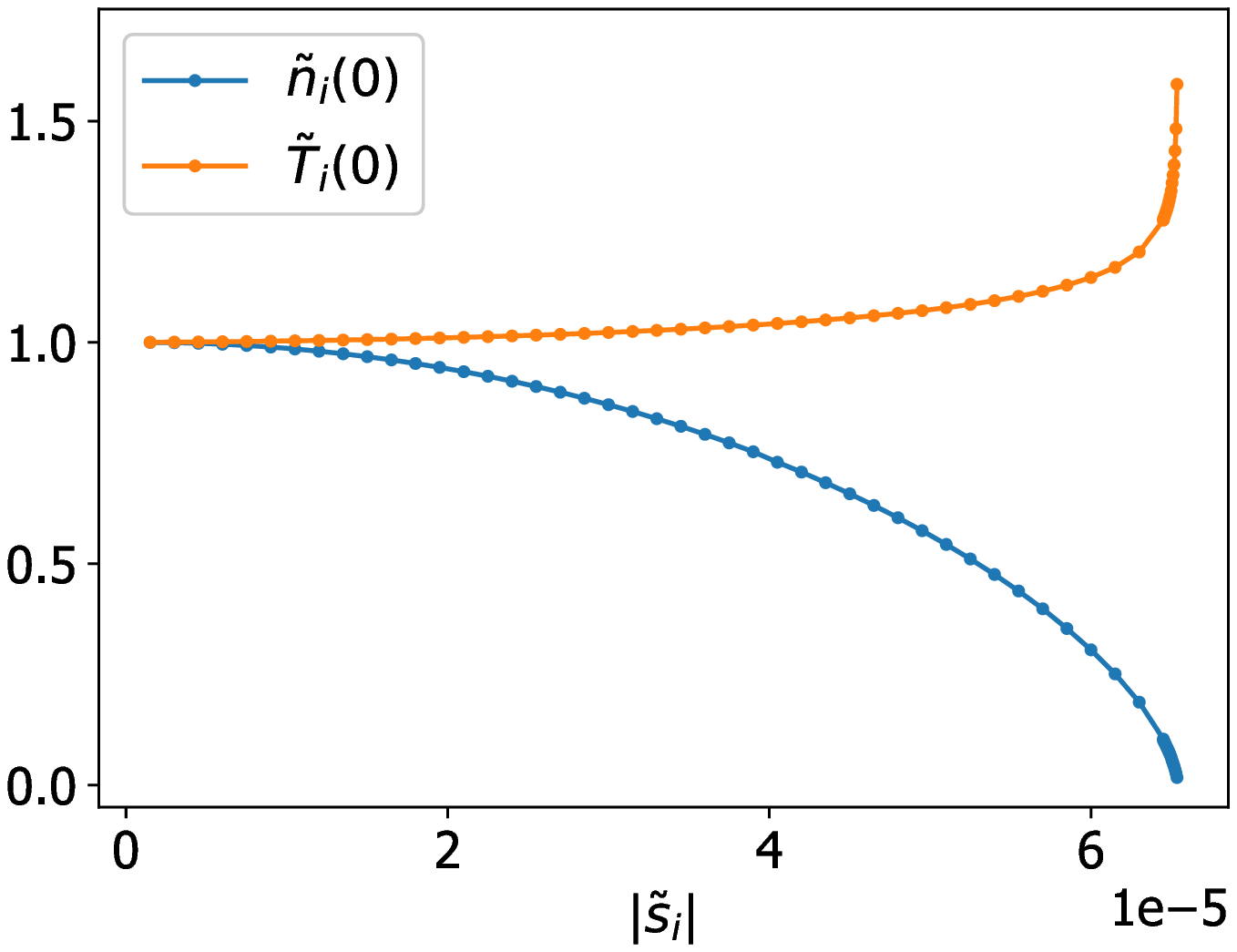}
    \caption{Values of angular velocity at the center of the cylinder and Mach number (maximal azimuthal velocity) (top), and values of density and temperature at the center of the cylinder, as a function of $|\ts_i|$ (bottom). These nonlinear calculations were performed using MITNS, with $\rho_*= 0.01$ and $\epsilon=0.1$.}
    \label{fig:param_scan}
\end{figure}

Even at Mach numbers in which a solution for the nonlinear case exists, the solution might develop an angular velocity boundary layer at $\tr = 0$ which might be nonphysical. This boundary layer enforces the axisymmetry condition $\tomega_i'(0)=0$, which appears in (\ref{eq:wcybc}). The boundary layer width, shrinks rapidly around $M=2$, while the temperature rises, as can be seen in Figure \ref{fig:param_scan}. When the boundary layer width becomes smaller than an ion Larmor radius, the solution becomes nonphysical. See appendix.
Steady state solutions for $M\gtrsim2$ don't exist in the full nonlinear case, as the viscosity drops quite significantly, and it could not balance the torque produced by the magnetic field.

Figure \ref{fig:param_scan} presents the values of the density, temperature, and angular velocity at the center of the cylinder, and the Mach number as a function of the particle sink magnitude. 

\begin{figure}
    \centering
    \includegraphics[width = \columnwidth]{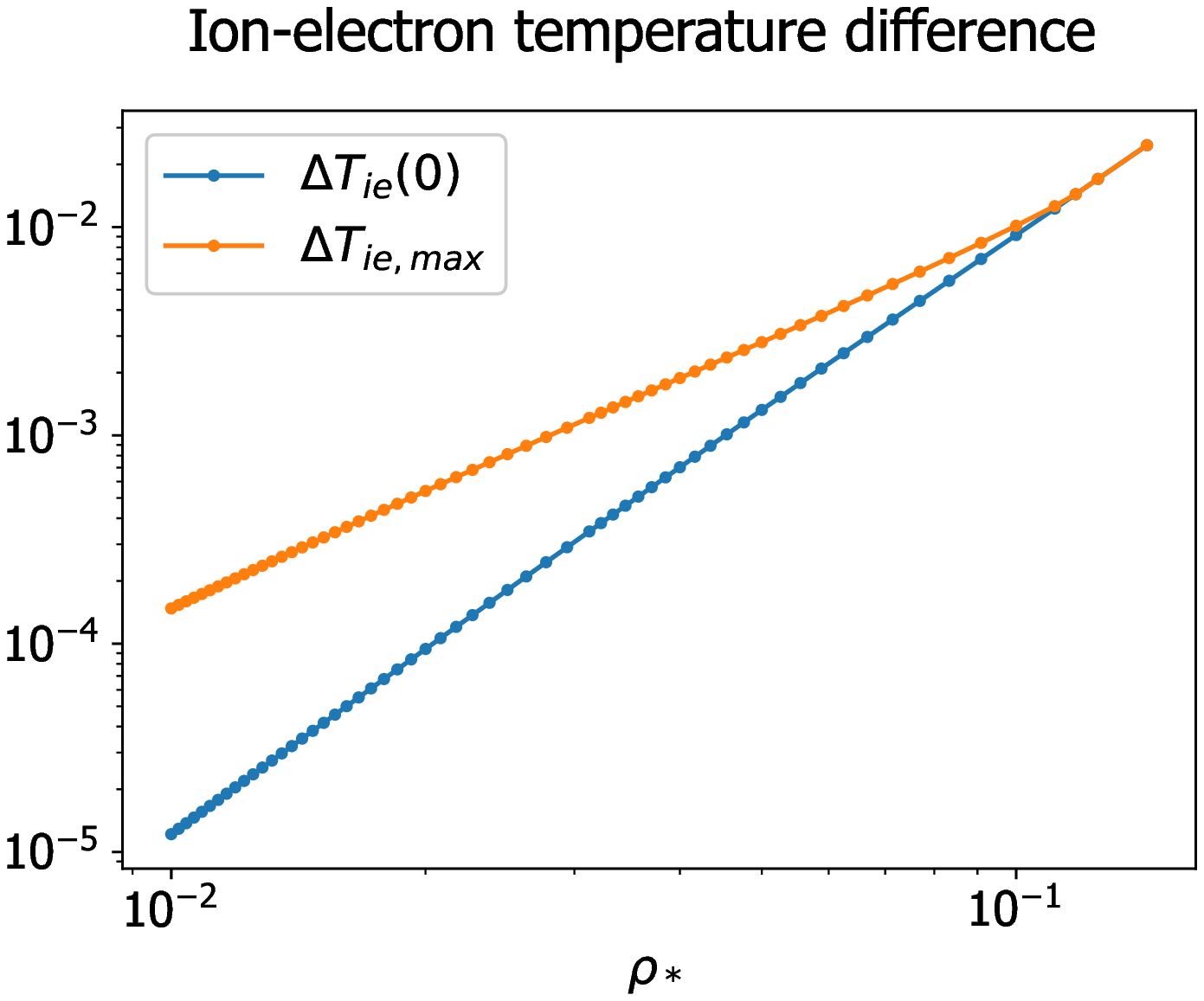}
    \includegraphics[width = \columnwidth]{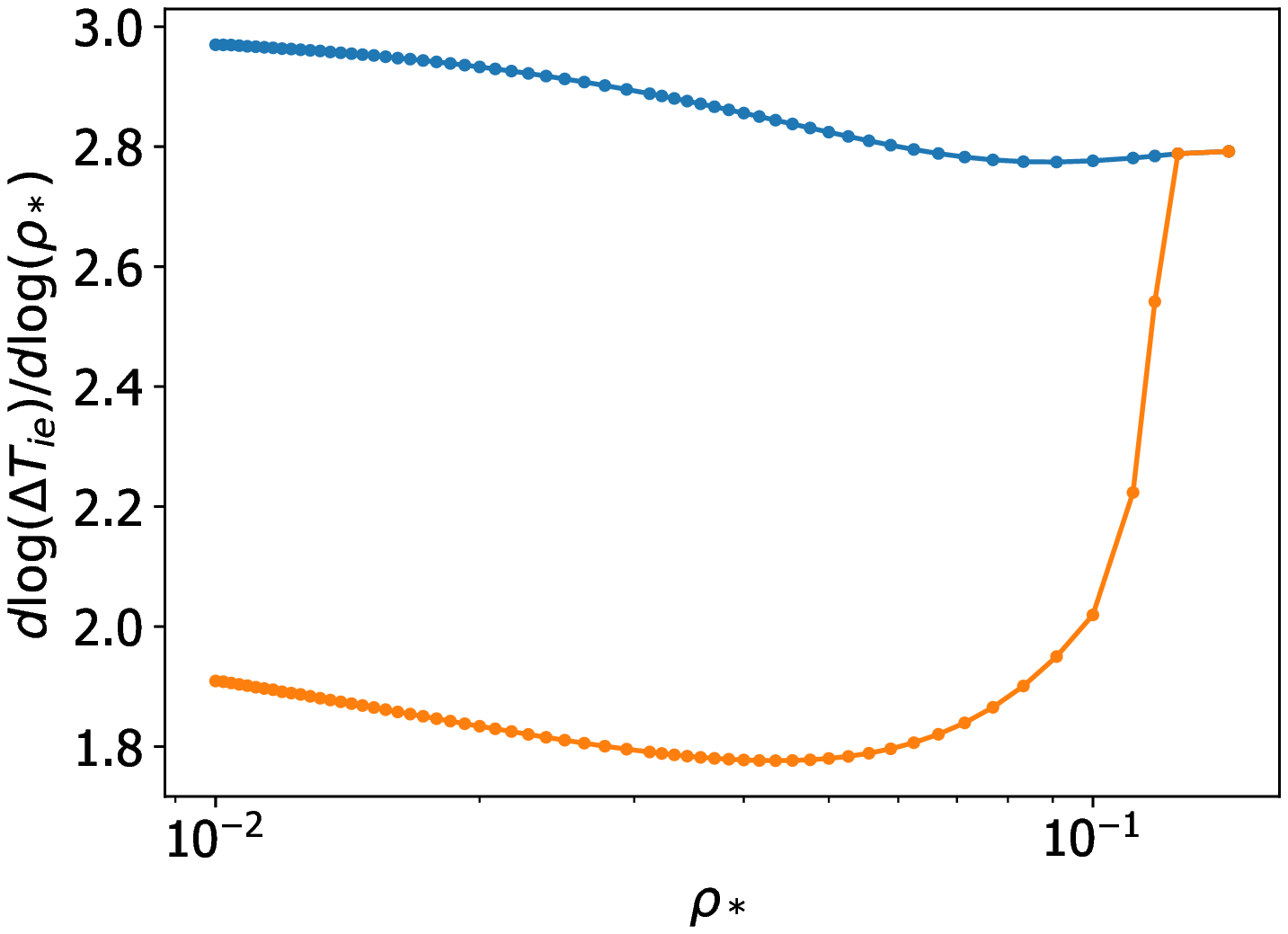}
    \caption{Temperature difference between ions and electrons at the center of the cylinder, and the maximal temperature difference as a function of $\rho_*$ (top), and the power-law dependence of the same curves (bottom). Results from the nonlinear solution in MITNS. The ion sink term was $\ts_i = -5 \rho_*^2\epsilon$, such that $F_i\sim\rho_*$, with $\epsilon=0.1$.}
    \label{fig:Delta_T}
\end{figure}

Figure \ref{fig:Delta_T} shows the temperature difference between ions and electrons at the center of the cylinder, and the maximal value of the temperature difference between the species, as a function of $\rho_*$. The sink term was chosen such that $F_i/\rho_*$ would remain constant at the boundary, and the azimuthal Mach number would approximately be constant at $M\approx 0.93$, with small variation due to the nonlinear nature of the system. This logarithmic plot shows a near-perfect power law, with the two curves crossing at $\rho_* = 0.11$, when the center of the cylinder becomes the point of maximal temperature difference between species. The maximal temperature difference depends very nearly on $\rho_*^2$, even in the nonlinear case.

\section{Conclusion}
A solution to the flow and temperature profiles in a highly magnetized rotating cylindrical two-fluid (ion-electron) plasma, driven by constant ion charge extraction following Braginskii's fluid was investigated.
First, a leading-order solution in the low-flow limit was presented.
Second, the physical validity of the asymptotic solution was considered, and the existence of a hot-ion mode was evaluated. It was shown that the collisional temperature coupling between the fluids is stronger than the difference in viscous heating between the fluids by a factor of $\sme/\rho_*^{2}$, and this limits the temperature difference to be of $\mathcal{O}(\rho_*^2/\sme)$.

The ratio of ion to electron heating is the same as the ratio of viscosity coefficients $\tm_e^{3/2}$. Picking a $\rho_*\sim \sqrt[4]{\tm_e} \sim 0.1$ would bring the temperature difference to $\mathcal{O}(0.1 - 1)$, depending on the exact sink term magnitude. This would be a large $\rho_*$, but not impossibly so - some magnetic traps or FRCs\cite{Steinhauer2011} operate in this regime.
Finally, the departure of the non-linear solution from the linear approximation at moderate Mach numbers ($M\gtrsim0.5$) was demonstrated and is explained by the hollowing-out of the viscosity and heat conductivity profiles. 

We have shown the inherent difficulties in achieving a significant hot-ion mode, even in the absence of electron heating, due to the limitation on ion heating. We have shown that low magnetization devices present an easier avenue for a hot-ion mode. Paradoxically, this is a point in favor of devices that contain fewer ion Larmor radii - which can be accomplished using a smaller magnetic field or a smaller device size - in view of the large advantage a hot-ion mode might present for a fusion plasma. 

There are other possible ion sink profiles, $\ts_i(\tr)$, and the details of such solutions may differ from the solution presented here. The main effect discussed here, however, the hollowing out of the density and the viscosity profiles, is weakly dependant on the specific $\ts_i(\tr)$ profile, and requires only the functional dependence $\tilde \eta_{1} \propto \tn^2$.

\subsection*{Acknowledgments}
This work was supported by Cornell NNSA 83228-10966 [Prime No. DOE (NNSA) DE-NA0003764] and by NSF-PHY-1805316.
IEO also acknowledges support from the DOE FES Fellowship [DOE Contract DE-SC0014664].
\section*{Author Declarations}
\subsection*{Conflict of interest}
The authors have no conflicts to disclose.
\subsection*{Data availability}
The data that support the findings of this study are available from the corresponding author upon reasonable request.

\appendix

\section{Boundary layers}~\label{Appendix:1}
The above solution includes several boundary layers, in the electron angular velocity, effective heat transfer coefficient, and the temperature difference between electrons and ions.
The arguments of the boundary layers are expressed in terms of the electron viscosity and heat transfer coefficients.  
Using Braginskii's classical transport coefficients, the argument of the electron angular velocity boundary layer term is
\begin{gather}
    \frac{\tr}{\rho_{*}}\sqrt{\frac{ \tm_e \tn_e \tilde \nu_{ei}}{\tilde \eta_{1e}}} = \frac{\tr}{\rho_{*}}\frac{ \tB}{\sqrt{0.51\tT_e\tm_e}} =1.4 \frac{r\Omega_e}{v_{the}} = 1.4\frac{r}{\rho_e}.
\end{gather}
Here, $\rho_e$ is the electron Larmor radius. This boundary layer is thinner than a Larmor orbit. As such, the classical transport model breaks down at these length scales, hence, we must abandon this boundary layer solution and the boundary condition that produced it, and resort to a full slip condition for the electron fluid at the outer edge of the cylinder.
This same boundary layer appears in the effective electron heat conduction, and must be discarded there as well.


As such, in the limit of Braginskii's classical transport, equations (\ref{eq:ewbc}), (\ref{eq:ewsol}), and (\ref{eq:kappaeeff}) should simply read
\begin{gather}
    (\tilde\nabla\cdot\tilde\pi_e)_\theta(1)=0,
    \\\tomega_e = \tomega_i-\frac{3}{2}\frac{\rho_{*}}{\tr \tB}\dv{\tT_e}{\tr},\\
    \tilde \kappa_{e_{eff}} = 6.91\frac{\tp_e\tilde \nu_{ei}}{\tm_e\tOmega_e^2}.
\end{gather}
However, some non-classical transport effects~\cite{rognlien1999}, such as turbulent anomalous transport, or the effects of perturbations to the magnetic field~\cite{Finn1992}, might enhance the electron viscosity above Braginskii's values, while still maintaining the validity of the fluid approximation. 
In these cases, if the effective electron viscosity might be large enough such that the boundary layer thickness might encompass several Larmor orbits, the boundary layer solution in the electron angular velocity might be a true physical effect.

The argument of the temperature-difference boundary layer, using Braginskii's classical transport coefficients, is 
\begin{gather}
    \frac{\tr}{\rho_*}\sqrt{3\tn_i \tilde \nu_{ie}\left(\tilde\kappa_{e_{eff}}^{-1}+\tilde\kappa_{i}^{-1}\right)}=\frac{\tr}{\rho_*}\frac{0.71}{\sqrt{\tm_i}}.
\end{gather}
This boundary layer does encompass many electron gyro-orbits, and is plausible because of it. 
Physically, it must be enforced by some effect on the boundary. We are not concerned with modeling the plasma-surface interactions that might lead to this electron temperature boundary layer, and conclude that the fluid approximation is valid for a boundary layer, with, possibly a different boundary value for the electrons other than the ions.

Note that the magnitude of $\Delta T_{ei}$, for the constant coefficient solution, is controlled by the scaling of the flux ($\ts_i$ or equivalently $F_i$), which directly affects $\tomega_i'$, and is not dependent on the exact form of the viscosity or heat transfer coefficients. Trying to maximize $\tT_i-\tT_e$ in equation (\ref{eq:DelTeisol}), by changing $\tilde \kappa_{e_{eff}}$ only, or in conjunction with $\tilde \eta_{1e}$, using an anomalous electron transport, for example, would yield limited results.

In a fusion reactor, when $\alpha$-channeling is missing or is insufficiently effective at removing the fusion ash, the ash would slow down on the electrons and introduce significant electron heating.  Indeed this mechanism typically causes a hot electron mode. Alternatively, Bremsstrahlung radiation introduces a heat sink in the electrons, which would help maintain a hot-ion mode, when dealing with mildly relativistic plasmas such as in $p-B^{11}$ reactors.

\section*{References}
\input{main.bbl}
\end{document}

%% file: Shortcuts.tex
\newcommand{\br}{\mathbf{r}}
\newcommand{\bv}{\mathbf{v}}

\newcommand{\bB}{\mathbf{B}}

\newcommand{\bj}{\mathbf{j}}
\newcommand{\bR}{\mathbf{R}}

\newcommand{\tr}{\tilde{r}}
\newcommand{\tOmega}{\tilde \Omega}
\newcommand{\tp}{\tilde{p}}

\newcommand{\tn}{\tilde{n}}
\newcommand{\tT}{\tilde{T}}

\newcommand{\tv}{\tilde{v}}

\newcommand{\tB}{\tilde{B}}

\newcommand{\tE}{\tilde{E}}

\newcommand{\ts}{\tilde{s}}
\newcommand{\tm}{\tilde{m}}
\newcommand{\pdv}[2]{\frac{\partial #1}{\partial #2}}
\newcommand{\dv}[2]{\frac{d #1}{d #2}}

\newcommand{\tomega}{\tilde{\omega}}
\newcommand{\sme}{\sqrt{\tm_e}}

%% file: Authors.tex
\author{T. Rubin}
\email{trubin@princeton.edu}
\affiliation{Department of Astrophysical Sciences, Princeton University, Princeton, New Jersey 08544, USA}

\author{E. J. Kolmes}
\affiliation{Department of Astrophysical Sciences, Princeton University, Princeton, New Jersey 08544, USA}
\author{I. E. Ochs}
\affiliation{Department of Astrophysical Sciences, Princeton University, Princeton, New Jersey 08544, USA}
\author{M. E. Mlodik}
\affiliation{Department of Astrophysical Sciences, Princeton University, Princeton, New Jersey 08544, USA}
\author{N. J. Fisch}
\affiliation{Department of Astrophysical Sciences, Princeton University, Princeton, New Jersey 08544, USA}

%% file: main.bbl
\providecommand{\noopsort}[1]{}\providecommand{\singleletter}[1]{#1}%

%% file: main.bbl
\begin{thebibliography}{36}%
\makeatletter
\providecommand \@ifxundefined [1]{%
 \@ifx{#1\undefined}
}%
\providecommand \@ifnum [1]{%
 \ifnum #1\expandafter \@firstoftwo
 \else \expandafter \@secondoftwo
 \fi
}%
\providecommand \@ifx [1]{%
 \ifx #1\expandafter \@firstoftwo
 \else \expandafter \@secondoftwo
 \fi
}%
\providecommand \natexlab [1]{#1}%
\providecommand \enquote  [1]{``#1''}%
\providecommand \bibnamefont  [1]{#1}%
\providecommand \bibfnamefont [1]{#1}%
\providecommand \citenamefont [1]{#1}%
\providecommand \href@noop [0]{\@secondoftwo}%
\providecommand \href [0]{\begingroup \@sanitize@url \@href}%
\providecommand \@href[1]{\@@startlink{#1}\@@href}%
\providecommand \@@href[1]{\endgroup#1\@@endlink}%
\providecommand \@sanitize@url [0]{\catcode `\\12\catcode `\$12\catcode
  `\&12\catcode `\#12\catcode `\^12\catcode `\_12\catcode `\%12\relax}%
\providecommand \@@startlink[1]{}%
\providecommand \@@endlink[0]{}%
\providecommand \url  [0]{\begingroup\@sanitize@url \@url }%
\providecommand \@url [1]{\endgroup\@href {#1}{\urlprefix }}%
\providecommand \urlprefix  [0]{URL }%
\providecommand \Eprint [0]{\href }%
\providecommand \doibase [0]{https://doi.org/}%
\providecommand \selectlanguage [0]{\@gobble}%
\providecommand \bibinfo  [0]{\@secondoftwo}%
\providecommand \bibfield  [0]{\@secondoftwo}%
\providecommand \translation [1]{[#1]}%
\providecommand \BibitemOpen [0]{}%
\providecommand \bibitemStop [0]{}%
\providecommand \bibitemNoStop [0]{.\EOS\space}%
\providecommand \EOS [0]{\spacefactor3000\relax}%
\providecommand \BibitemShut  [1]{\csname bibitem#1\endcsname}%
\let\auto@bib@innerbib\@empty
\bibitem [{\citenamefont {Lehnert}(1971)}]{Lehnert1971}%
  \BibitemOpen
  \bibfield  {author} {\bibinfo {author} {\bibfnamefont {B.}~\bibnamefont
  {Lehnert}},\ }\href {https://doi.org/10.1088/0029-5515/11/5/010} {\bibfield
  {journal} {\bibinfo  {journal} {Nucl. Fusion}\ }\textbf {\bibinfo {volume}
  {11}},\ \bibinfo {pages} {485} (\bibinfo {year} {1971})}\BibitemShut
  {NoStop}%
\bibitem [{\citenamefont {Bekhtenev}\ \emph {et~al.}(1980)\citenamefont
  {Bekhtenev}, \citenamefont {Volosov}, \citenamefont
  {Pal{\textquotesingle}chikov}, \citenamefont {Pekker},\ and\ \citenamefont
  {Yudin}}]{Bekhtenev1980}%
  \BibitemOpen
  \bibfield  {author} {\bibinfo {author} {\bibfnamefont {A.}~\bibnamefont
  {Bekhtenev}}, \bibinfo {author} {\bibfnamefont {V.}~\bibnamefont {Volosov}},
  \bibinfo {author} {\bibfnamefont {V.}~\bibnamefont
  {Pal{\textquotesingle}chikov}}, \bibinfo {author} {\bibfnamefont
  {M.}~\bibnamefont {Pekker}},\ and\ \bibinfo {author} {\bibfnamefont
  {Y.}~\bibnamefont {Yudin}},\ }\href
  {https://doi.org/10.1088/0029-5515/20/5/007} {\bibfield  {journal} {\bibinfo
  {journal} {Nuclear Fusion}\ }\textbf {\bibinfo {volume} {20}},\ \bibinfo
  {pages} {579} (\bibinfo {year} {1980})}\BibitemShut {NoStop}%
\bibitem [{\citenamefont {Hassam}(1999)}]{Hassam1999}%
  \BibitemOpen
  \bibfield  {author} {\bibinfo {author} {\bibfnamefont {A.~B.}\ \bibnamefont
  {Hassam}},\ }\href {https://doi.org/10.1063/1.873636} {\bibfield  {journal}
  {\bibinfo  {journal} {Physics of Plasmas}\ }\textbf {\bibinfo {volume} {6}},\
  \bibinfo {pages} {3738} (\bibinfo {year} {1999})}\BibitemShut {NoStop}%
\bibitem [{\citenamefont {Rax}\ \emph {et~al.}(2017)\citenamefont {Rax},
  \citenamefont {Gueroult},\ and\ \citenamefont {Fisch}}]{Rax2017}%
  \BibitemOpen
  \bibfield  {author} {\bibinfo {author} {\bibfnamefont {J.-M.}\ \bibnamefont
  {Rax}}, \bibinfo {author} {\bibfnamefont {R.}~\bibnamefont {Gueroult}},\ and\
  \bibinfo {author} {\bibfnamefont {N.~J.}\ \bibnamefont {Fisch}},\ }\href
  {https://doi.org/10.1063/1.4977919} {\bibfield  {journal} {\bibinfo
  {journal} {Phys. Plasmas}\ }\textbf {\bibinfo {volume} {24}},\ \bibinfo
  {pages} {032504} (\bibinfo {year} {2017})}\BibitemShut {NoStop}%
\bibitem [{\citenamefont {Ochs}\ and\ \citenamefont
  {Fisch}(2017)}]{Ochs2017ii}%
  \BibitemOpen
  \bibfield  {author} {\bibinfo {author} {\bibfnamefont {I.~E.}\ \bibnamefont
  {Ochs}}\ and\ \bibinfo {author} {\bibfnamefont {N.~J.}\ \bibnamefont
  {Fisch}},\ }\href {https://doi.org/10.1063/1.4991510} {\bibfield  {journal}
  {\bibinfo  {journal} {Phys. Plasmas}\ }\textbf {\bibinfo {volume} {24}},\
  \bibinfo {pages} {092513} (\bibinfo {year} {2017})}\BibitemShut {NoStop}%
\bibitem [{\citenamefont {Ellis}\ \emph {et~al.}(2001)\citenamefont {Ellis},
  \citenamefont {Hassam}, \citenamefont {Messer},\ and\ \citenamefont
  {Osborn}}]{Ellis2001}%
  \BibitemOpen
  \bibfield  {author} {\bibinfo {author} {\bibfnamefont {R.~F.}\ \bibnamefont
  {Ellis}}, \bibinfo {author} {\bibfnamefont {A.~B.}\ \bibnamefont {Hassam}},
  \bibinfo {author} {\bibfnamefont {S.}~\bibnamefont {Messer}},\ and\ \bibinfo
  {author} {\bibfnamefont {B.~R.}\ \bibnamefont {Osborn}},\ }\href
  {https://doi.org/10.1063/1.1350957} {\bibfield  {journal} {\bibinfo
  {journal} {Physics of Plasmas}\ }\textbf {\bibinfo {volume} {8}},\ \bibinfo
  {pages} {2057} (\bibinfo {year} {2001})}\BibitemShut {NoStop}%
\bibitem [{\citenamefont {Ellis}\ \emph {et~al.}(2005)\citenamefont {Ellis},
  \citenamefont {Case}, \citenamefont {Elton}, \citenamefont {Ghosh},
  \citenamefont {Griem}, \citenamefont {Hassam}, \citenamefont {Lunsford},
  \citenamefont {Messer},\ and\ \citenamefont {Teodorescu}}]{Ellis2005}%
  \BibitemOpen
  \bibfield  {author} {\bibinfo {author} {\bibfnamefont {R.~F.}\ \bibnamefont
  {Ellis}}, \bibinfo {author} {\bibfnamefont {A.}~\bibnamefont {Case}},
  \bibinfo {author} {\bibfnamefont {R.}~\bibnamefont {Elton}}, \bibinfo
  {author} {\bibfnamefont {J.}~\bibnamefont {Ghosh}}, \bibinfo {author}
  {\bibfnamefont {H.}~\bibnamefont {Griem}}, \bibinfo {author} {\bibfnamefont
  {A.}~\bibnamefont {Hassam}}, \bibinfo {author} {\bibfnamefont
  {R.}~\bibnamefont {Lunsford}}, \bibinfo {author} {\bibfnamefont
  {S.}~\bibnamefont {Messer}},\ and\ \bibinfo {author} {\bibfnamefont
  {C.}~\bibnamefont {Teodorescu}},\ }\href {https://doi.org/10.1063/1.1896954}
  {\bibfield  {journal} {\bibinfo  {journal} {Phys. Plasmas}\ }\textbf
  {\bibinfo {volume} {12}},\ \bibinfo {pages} {055704} (\bibinfo {year}
  {2005})}\BibitemShut {NoStop}%
\bibitem [{\citenamefont {Teodorescu}\ \emph {et~al.}(2010)\citenamefont
  {Teodorescu}, \citenamefont {Young}, \citenamefont {Swan}, \citenamefont
  {Ellis}, \citenamefont {Hassam},\ and\ \citenamefont
  {Romero-Talamas}}]{Teodorescu2010}%
  \BibitemOpen
  \bibfield  {author} {\bibinfo {author} {\bibfnamefont {C.}~\bibnamefont
  {Teodorescu}}, \bibinfo {author} {\bibfnamefont {W.~C.}\ \bibnamefont
  {Young}}, \bibinfo {author} {\bibfnamefont {G.~W.~S.}\ \bibnamefont {Swan}},
  \bibinfo {author} {\bibfnamefont {R.~F.}\ \bibnamefont {Ellis}}, \bibinfo
  {author} {\bibfnamefont {A.~B.}\ \bibnamefont {Hassam}},\ and\ \bibinfo
  {author} {\bibfnamefont {C.~A.}\ \bibnamefont {Romero-Talamas}},\ }\href
  {https://doi.org/10.1103/PhysRevLett.105.085003} {\bibfield  {journal}
  {\bibinfo  {journal} {Phys. Rev. Lett.}\ }\textbf {\bibinfo {volume} {105}},\
  \bibinfo {pages} {085003} (\bibinfo {year} {2010})}\BibitemShut {NoStop}%
\bibitem [{\citenamefont {Rax}\ and\ \citenamefont {Gueroult}(2016)}]{Rax2016}%
  \BibitemOpen
  \bibfield  {author} {\bibinfo {author} {\bibfnamefont {J.-M.}\ \bibnamefont
  {Rax}}\ and\ \bibinfo {author} {\bibfnamefont {R.}~\bibnamefont {Gueroult}},\
  }\href {https://doi.org/10.1017/S0022377816000878} {\bibfield  {journal}
  {\bibinfo  {journal} {{J. Plasma Phys.}}\ }\textbf {\bibinfo {volume}
  {{82}}},\ \bibinfo {pages} {{595820504}} (\bibinfo {year}
  {{2016}})}\BibitemShut {NoStop}%
\bibitem [{\citenamefont {Zweben}\ \emph {et~al.}(2018)\citenamefont {Zweben},
  \citenamefont {Gueroult},\ and\ \citenamefont {Fisch}}]{Zweben2018}%
  \BibitemOpen
  \bibfield  {author} {\bibinfo {author} {\bibfnamefont {S.~J.}\ \bibnamefont
  {Zweben}}, \bibinfo {author} {\bibfnamefont {R.}~\bibnamefont {Gueroult}},\
  and\ \bibinfo {author} {\bibfnamefont {N.~J.}\ \bibnamefont {Fisch}},\ }\href
  {https://doi.org/10.1063/1.5042845} {\bibfield  {journal} {\bibinfo
  {journal} {Phys. Plasmas}\ }\textbf {\bibinfo {volume} {25}},\ \bibinfo
  {pages} {090901} (\bibinfo {year} {2018})}\BibitemShut {NoStop}%
\bibitem [{\citenamefont {Ochs}\ \emph {et~al.}(2017)\citenamefont {Ochs},
  \citenamefont {Gueroult}, \citenamefont {Fisch},\ and\ \citenamefont
  {Zweben}}]{Ochs2017iii}%
  \BibitemOpen
  \bibfield  {author} {\bibinfo {author} {\bibfnamefont {I.~E.}\ \bibnamefont
  {Ochs}}, \bibinfo {author} {\bibfnamefont {R.}~\bibnamefont {Gueroult}},
  \bibinfo {author} {\bibfnamefont {N.~J.}\ \bibnamefont {Fisch}},\ and\
  \bibinfo {author} {\bibfnamefont {S.~J.}\ \bibnamefont {Zweben}},\ }\href
  {https://doi.org/10.1063/1.4978949} {\bibfield  {journal} {\bibinfo
  {journal} {Phys. Plasmas}\ }\textbf {\bibinfo {volume} {24}},\ \bibinfo
  {pages} {043503} (\bibinfo {year} {2017})}\BibitemShut {NoStop}%
\bibitem [{\citenamefont {Gueroult}\ and\ \citenamefont
  {Fisch}(2012)}]{Gueroult2012MCMF}%
  \BibitemOpen
  \bibfield  {author} {\bibinfo {author} {\bibfnamefont {R.}~\bibnamefont
  {Gueroult}}\ and\ \bibinfo {author} {\bibfnamefont {N.~J.}\ \bibnamefont
  {Fisch}},\ }\href {https://doi.org/10.1063/1.4771674} {\bibfield  {journal}
  {\bibinfo  {journal} {Phys. Plasmas}\ }\textbf {\bibinfo {volume} {19}},\
  \bibinfo {pages} {122503} (\bibinfo {year} {2012})}\BibitemShut {NoStop}%
\bibitem [{\citenamefont {Gueroult}\ \emph {et~al.}(2015)\citenamefont
  {Gueroult}, \citenamefont {Hobbs},\ and\ \citenamefont
  {Fisch}}]{Gueroult2015}%
  \BibitemOpen
  \bibfield  {author} {\bibinfo {author} {\bibfnamefont {R.}~\bibnamefont
  {Gueroult}}, \bibinfo {author} {\bibfnamefont {D.~T.}\ \bibnamefont
  {Hobbs}},\ and\ \bibinfo {author} {\bibfnamefont {N.~J.}\ \bibnamefont
  {Fisch}},\ }\href {https://doi.org/10.1016/j.jhazmat.2015.04.058} {\bibfield
  {journal} {\bibinfo  {journal} {J. Hazard. Mater.}\ }\textbf {\bibinfo
  {volume} {297}},\ \bibinfo {pages} {153} (\bibinfo {year}
  {2015})}\BibitemShut {NoStop}%
\bibitem [{\citenamefont {Gueroult}\ \emph {et~al.}(2019)\citenamefont
  {Gueroult}, \citenamefont {Rax},\ and\ \citenamefont
  {Fisch}}]{Gueroult2019ii}%
  \BibitemOpen
  \bibfield  {author} {\bibinfo {author} {\bibfnamefont {R.}~\bibnamefont
  {Gueroult}}, \bibinfo {author} {\bibfnamefont {J.-M.}\ \bibnamefont {Rax}},\
  and\ \bibinfo {author} {\bibfnamefont {N.~J.}\ \bibnamefont {Fisch}},\ }\href
  {https://doi.org/10.1063/1.5126083} {\bibfield  {journal} {\bibinfo
  {journal} {Phys. Plasmas}\ }\textbf {\bibinfo {volume} {26}},\ \bibinfo
  {pages} {122106} (\bibinfo {year} {2019})}\BibitemShut {NoStop}%
\bibitem [{\citenamefont {Fetterman}\ and\ \citenamefont
  {Fisch}(2011)}]{Fetterman2011b}%
  \BibitemOpen
  \bibfield  {author} {\bibinfo {author} {\bibfnamefont {A.~J.}\ \bibnamefont
  {Fetterman}}\ and\ \bibinfo {author} {\bibfnamefont {N.~J.}\ \bibnamefont
  {Fisch}},\ }\href {https://doi.org/10.1063/1.3631793} {\bibfield  {journal}
  {\bibinfo  {journal} {Phys. Plasmas}\ }\textbf {\bibinfo {volume} {18}},\
  \bibinfo {pages} {094503} (\bibinfo {year} {2011})}\BibitemShut {NoStop}%
\bibitem [{\citenamefont {Bonnevier}(1966)}]{Bonnevier1966}%
  \BibitemOpen
  \bibfield  {author} {\bibinfo {author} {\bibfnamefont {B.}~\bibnamefont
  {Bonnevier}},\ }\href@noop {} {\bibfield  {journal} {\bibinfo  {journal}
  {Ark. Fys.}\ }\textbf {\bibinfo {volume} {33}},\ \bibinfo {pages} {255}
  (\bibinfo {year} {1966})}\BibitemShut {NoStop}%
\bibitem [{\citenamefont {Bonnevier}(1971)}]{Bonnevier1971}%
  \BibitemOpen
  \bibfield  {author} {\bibinfo {author} {\bibfnamefont {B.}~\bibnamefont
  {Bonnevier}},\ }\href {https://doi.org/10.1088/0032-1028/13/9/007} {\bibfield
   {journal} {\bibinfo  {journal} {Plasma Phys.}\ }\textbf {\bibinfo {volume}
  {13}},\ \bibinfo {pages} {763} (\bibinfo {year} {1971})}\BibitemShut
  {NoStop}%
\bibitem [{\citenamefont {Kolmes}\ \emph {et~al.}(2019)\citenamefont {Kolmes},
  \citenamefont {Ochs}, \citenamefont {Mlodik}, \citenamefont {Rax},
  \citenamefont {Gueroult},\ and\ \citenamefont {Fisch}}]{Kolmes2019}%
  \BibitemOpen
  \bibfield  {author} {\bibinfo {author} {\bibfnamefont {E.~J.}\ \bibnamefont
  {Kolmes}}, \bibinfo {author} {\bibfnamefont {I.~E.}\ \bibnamefont {Ochs}},
  \bibinfo {author} {\bibfnamefont {M.~E.}\ \bibnamefont {Mlodik}}, \bibinfo
  {author} {\bibfnamefont {J.-M.}\ \bibnamefont {Rax}}, \bibinfo {author}
  {\bibfnamefont {R.}~\bibnamefont {Gueroult}},\ and\ \bibinfo {author}
  {\bibfnamefont {N.~J.}\ \bibnamefont {Fisch}},\ }\href
  {https://doi.org/10.1063/1.5115788} {\bibfield  {journal} {\bibinfo
  {journal} {Physics of Plasmas}\ }\textbf {\bibinfo {volume} {26}},\ \bibinfo
  {pages} {082309} (\bibinfo {year} {2019})}\BibitemShut {NoStop}%
\bibitem [{\citenamefont {Rax}\ \emph {et~al.}(2019)\citenamefont {Rax},
  \citenamefont {Kolmes}, \citenamefont {Ochs}, \citenamefont {Fisch},\ and\
  \citenamefont {Gueroult}}]{Rax2019}%
  \BibitemOpen
  \bibfield  {author} {\bibinfo {author} {\bibfnamefont {J.-M.}\ \bibnamefont
  {Rax}}, \bibinfo {author} {\bibfnamefont {E.~J.}\ \bibnamefont {Kolmes}},
  \bibinfo {author} {\bibfnamefont {I.~E.}\ \bibnamefont {Ochs}}, \bibinfo
  {author} {\bibfnamefont {N.~J.}\ \bibnamefont {Fisch}},\ and\ \bibinfo
  {author} {\bibfnamefont {R.}~\bibnamefont {Gueroult}},\ }\href
  {https://doi.org/10.1063/1.5064520} {\bibfield  {journal} {\bibinfo
  {journal} {Phys. Plasmas}\ }\textbf {\bibinfo {volume} {26}},\ \bibinfo
  {pages} {012303} (\bibinfo {year} {2019})}\BibitemShut {NoStop}%
\bibitem [{\citenamefont {Rubin}\ \emph {et~al.}(2021)\citenamefont {Rubin},
  \citenamefont {Kolmes}, \citenamefont {Ochs}, \citenamefont {Mlodik},\ and\
  \citenamefont {Fisch}}]{Rubin2021}%
  \BibitemOpen
  \bibfield  {author} {\bibinfo {author} {\bibfnamefont {T.}~\bibnamefont
  {Rubin}}, \bibinfo {author} {\bibfnamefont {E.~J.}\ \bibnamefont {Kolmes}},
  \bibinfo {author} {\bibfnamefont {I.~E.}\ \bibnamefont {Ochs}}, \bibinfo
  {author} {\bibfnamefont {M.~E.}\ \bibnamefont {Mlodik}},\ and\ \bibinfo
  {author} {\bibfnamefont {N.~J.}\ \bibnamefont {Fisch}},\ }\href
  {https://doi.org/10.1063/5.0070292} {\bibfield  {journal} {\bibinfo
  {journal} {Physics of Plasmas}\ }\textbf {\bibinfo {volume} {28}},\ \bibinfo
  {pages} {122303} (\bibinfo {year} {2021})}\BibitemShut {NoStop}%
\bibitem [{\citenamefont {Kolmes}\ \emph
  {et~al.}(2021{\natexlab{a}})\citenamefont {Kolmes}, \citenamefont {Ochs},\
  and\ \citenamefont {Fisch}}]{Kolmes2021}%
  \BibitemOpen
  \bibfield  {author} {\bibinfo {author} {\bibfnamefont {E.~J.}\ \bibnamefont
  {Kolmes}}, \bibinfo {author} {\bibfnamefont {I.~E.}\ \bibnamefont {Ochs}},\
  and\ \bibinfo {author} {\bibfnamefont {N.~J.}\ \bibnamefont {Fisch}},\ }\href
  {https://doi.org/10.1016/j.cpc.2020.107511} {\bibfield  {journal} {\bibinfo
  {journal} {Comp. Phys. Communications}\ }\textbf {\bibinfo {volume} {258}},\
  \bibinfo {pages} {107511} (\bibinfo {year} {2021}{\natexlab{a}})}\BibitemShut
  {NoStop}%
\bibitem [{\citenamefont {Lawson}(1957)}]{Lawson1957}%
  \BibitemOpen
  \bibfield  {author} {\bibinfo {author} {\bibfnamefont {J.~D.}\ \bibnamefont
  {Lawson}},\ }\href {https://doi.org/10.1088/0370-1301/70/1/303} {\bibfield
  {journal} {\bibinfo  {journal} {Proceedings of the Physical Society. Section
  B}\ }\textbf {\bibinfo {volume} {70}},\ \bibinfo {pages} {6} (\bibinfo {year}
  {1957})}\BibitemShut {NoStop}%
\bibitem [{\citenamefont {Clarke}(1980)}]{Clarke1980}%
  \BibitemOpen
  \bibfield  {author} {\bibinfo {author} {\bibfnamefont {J.}~\bibnamefont
  {Clarke}},\ }\href {https://doi.org/10.1088/0029-5515/20/5/005} {\bibfield
  {journal} {\bibinfo  {journal} {Nuclear Fusion}\ }\textbf {\bibinfo {volume}
  {20}},\ \bibinfo {pages} {563} (\bibinfo {year} {1980})}\BibitemShut
  {NoStop}%
\bibitem [{\citenamefont {Kolmes}\ \emph
  {et~al.}(2021{\natexlab{b}})\citenamefont {Kolmes}, \citenamefont {Ochs},
  \citenamefont {Mlodik},\ and\ \citenamefont {Fisch}}]{Kolmes2021ii}%
  \BibitemOpen
  \bibfield  {author} {\bibinfo {author} {\bibfnamefont {E.~J.}\ \bibnamefont
  {Kolmes}}, \bibinfo {author} {\bibfnamefont {I.~E.}\ \bibnamefont {Ochs}},
  \bibinfo {author} {\bibfnamefont {M.~E.}\ \bibnamefont {Mlodik}},\ and\
  \bibinfo {author} {\bibfnamefont {N.~J.}\ \bibnamefont {Fisch}},\ }\href
  {https://doi.org/10.1103/PhysRevE.104.015209} {\bibfield  {journal} {\bibinfo
   {journal} {Phys. Rev. E}\ }\textbf {\bibinfo {volume} {104}},\ \bibinfo
  {pages} {015209} (\bibinfo {year} {2021}{\natexlab{b}})}\BibitemShut
  {NoStop}%
\bibitem [{\citenamefont {Fisch}\ and\ \citenamefont {Rax}(1992)}]{Fisch1992}%
  \BibitemOpen
  \bibfield  {author} {\bibinfo {author} {\bibfnamefont {N.~J.}\ \bibnamefont
  {Fisch}}\ and\ \bibinfo {author} {\bibfnamefont {J.-M.}\ \bibnamefont
  {Rax}},\ }\href {https://doi.org/10.1103/PhysRevLett.69.612} {\bibfield
  {journal} {\bibinfo  {journal} {Phys. Rev. Lett.}\ }\textbf {\bibinfo
  {volume} {69}},\ \bibinfo {pages} {612} (\bibinfo {year} {1992})}\BibitemShut
  {NoStop}%
\bibitem [{\citenamefont {Fisch}\ and\ \citenamefont
  {Herrmann}(1994)}]{Fisch1994}%
  \BibitemOpen
  \bibfield  {author} {\bibinfo {author} {\bibfnamefont {N.~J.}\ \bibnamefont
  {Fisch}}\ and\ \bibinfo {author} {\bibfnamefont {M.~C.}\ \bibnamefont
  {Herrmann}},\ }\href {https://doi.org/10.1088/0029-5515/34/12/I01} {\bibfield
   {journal} {\bibinfo  {journal} {Nucl. Fusion}\ }\textbf {\bibinfo {volume}
  {34}},\ \bibinfo {pages} {1541} (\bibinfo {year} {1994})}\BibitemShut
  {NoStop}%
\bibitem [{\citenamefont {Braginskii}(1965)}]{Braginskii1965}%
  \BibitemOpen
  \bibfield  {author} {\bibinfo {author} {\bibfnamefont {S.~I.}\ \bibnamefont
  {Braginskii}},\ }\bibinfo {title} {Transport processes in a plasma},\ in\
  \href@noop {} {\emph {\bibinfo {booktitle} {Reviews of Plasma Physics}}},\
  Vol.~\bibinfo {volume} {1},\ \bibinfo {editor} {edited by\ \bibinfo {editor}
  {\bibfnamefont {M.~A.}\ \bibnamefont {Leontovich}}}\ (\bibinfo  {publisher}
  {Consultants Bureau},\ \bibinfo {address} {New York},\ \bibinfo {year}
  {1965})\ p.\ \bibinfo {pages} {205}\BibitemShut {NoStop}%
\bibitem [{\citenamefont {Epperlein}\ and\ \citenamefont
  {Haines}(1986)}]{Epperlein1986}%
  \BibitemOpen
  \bibfield  {author} {\bibinfo {author} {\bibfnamefont {E.~M.}\ \bibnamefont
  {Epperlein}}\ and\ \bibinfo {author} {\bibfnamefont {M.~G.}\ \bibnamefont
  {Haines}},\ }\href {https://doi.org/10.1063/1.865901} {\bibfield  {journal}
  {\bibinfo  {journal} {Phys. Fluids}\ }\textbf {\bibinfo {volume} {29}},\
  \bibinfo {pages} {1029} (\bibinfo {year} {1986})}\BibitemShut {NoStop}%
\bibitem [{\citenamefont {Simakov}(2022)}]{Simakov2022}%
  \BibitemOpen
  \bibfield  {author} {\bibinfo {author} {\bibfnamefont {A.~N.}\ \bibnamefont
  {Simakov}},\ }\href {https://doi.org/10.1063/5.0080151} {\bibfield  {journal}
  {\bibinfo  {journal} {Physics of Plasmas}\ }\textbf {\bibinfo {volume}
  {29}},\ \bibinfo {pages} {022304} (\bibinfo {year} {2022})}\BibitemShut
  {NoStop}%
\bibitem [{\citenamefont {Ji}\ and\ \citenamefont
  {Held}(2013)}]{Jeong-Young2013}%
  \BibitemOpen
  \bibfield  {author} {\bibinfo {author} {\bibfnamefont {J.-Y.}\ \bibnamefont
  {Ji}}\ and\ \bibinfo {author} {\bibfnamefont {E.~D.}\ \bibnamefont {Held}},\
  }\href {https://doi.org/10.1063/1.4801022} {\bibfield  {journal} {\bibinfo
  {journal} {Physics of Plasmas}\ }\textbf {\bibinfo {volume} {20}},\ \bibinfo
  {pages} {042114} (\bibinfo {year} {2013})}\BibitemShut {NoStop}%
\bibitem [{\citenamefont {Sadler}\ \emph {et~al.}(2021)\citenamefont {Sadler},
  \citenamefont {Walsh},\ and\ \citenamefont {Li}}]{Sadler2021}%
  \BibitemOpen
  \bibfield  {author} {\bibinfo {author} {\bibfnamefont {J.~D.}\ \bibnamefont
  {Sadler}}, \bibinfo {author} {\bibfnamefont {C.~A.}\ \bibnamefont {Walsh}},\
  and\ \bibinfo {author} {\bibfnamefont {H.}~\bibnamefont {Li}},\ }\href
  {https://doi.org/10.1103/PhysRevLett.126.075001} {\bibfield  {journal}
  {\bibinfo  {journal} {Phys. Rev. Lett.}\ }\textbf {\bibinfo {volume} {126}},\
  \bibinfo {pages} {075001} (\bibinfo {year} {2021})}\BibitemShut {NoStop}%
\bibitem [{\citenamefont {Davies}\ \emph {et~al.}(2021)\citenamefont {Davies},
  \citenamefont {Wen}, \citenamefont {Ji},\ and\ \citenamefont
  {Held}}]{Davies2021}%
  \BibitemOpen
  \bibfield  {author} {\bibinfo {author} {\bibfnamefont {J.~R.}\ \bibnamefont
  {Davies}}, \bibinfo {author} {\bibfnamefont {H.}~\bibnamefont {Wen}},
  \bibinfo {author} {\bibfnamefont {J.-Y.}\ \bibnamefont {Ji}},\ and\ \bibinfo
  {author} {\bibfnamefont {E.~D.}\ \bibnamefont {Held}},\ }\href
  {https://doi.org/10.1063/5.0023445} {\bibfield  {journal} {\bibinfo
  {journal} {Physics of Plasmas}\ }\textbf {\bibinfo {volume} {28}},\ \bibinfo
  {pages} {012305} (\bibinfo {year} {2021})}\BibitemShut {NoStop}%
\bibitem [{\citenamefont {Fundamenski}\ and\ \citenamefont
  {Garcia}(2007)}]{Fundamenski2007}%
  \BibitemOpen
  \bibfield  {author} {\bibinfo {author} {\bibfnamefont {W.}~\bibnamefont
  {Fundamenski}}\ and\ \bibinfo {author} {\bibfnamefont {O.}~\bibnamefont
  {Garcia}},\ }\href@noop {} {\bibfield  {journal} {\bibinfo  {journal} {Report
  No. EFDA-JET-R (07) 01}\ } (\bibinfo {year} {2007})}\BibitemShut {NoStop}%
\bibitem [{\citenamefont {Steinhauer}(2011)}]{Steinhauer2011}%
  \BibitemOpen
  \bibfield  {author} {\bibinfo {author} {\bibfnamefont {L.~C.}\ \bibnamefont
  {Steinhauer}},\ }\href {https://doi.org/10.1063/1.3613680} {\bibfield
  {journal} {\bibinfo  {journal} {Physics of Plasmas}\ }\textbf {\bibinfo
  {volume} {18}},\ \bibinfo {pages} {070501} (\bibinfo {year}
  {2011})}\BibitemShut {NoStop}%
\bibitem [{\citenamefont {Rognlien}\ and\ \citenamefont
  {Ryutov}(1999)}]{rognlien1999}%
  \BibitemOpen
  \bibfield  {author} {\bibinfo {author} {\bibfnamefont {T.}~\bibnamefont
  {Rognlien}}\ and\ \bibinfo {author} {\bibfnamefont {D.}~\bibnamefont
  {Ryutov}},\ }\href@noop {} {\bibfield  {journal} {\bibinfo  {journal} {Plasma
  Physics Reports}\ }\textbf {\bibinfo {volume} {25}},\ \bibinfo {pages} {943}
  (\bibinfo {year} {1999})}\BibitemShut {NoStop}%
\bibitem [{\citenamefont {Finn}\ \emph {et~al.}(1992)\citenamefont {Finn},
  \citenamefont {Guzdar},\ and\ \citenamefont {Chernikov}}]{Finn1992}%
  \BibitemOpen
  \bibfield  {author} {\bibinfo {author} {\bibfnamefont {J.~M.}\ \bibnamefont
  {Finn}}, \bibinfo {author} {\bibfnamefont {P.~N.}\ \bibnamefont {Guzdar}},\
  and\ \bibinfo {author} {\bibfnamefont {A.~A.}\ \bibnamefont {Chernikov}},\
  }\href {https://doi.org/10.1063/1.860123} {\bibfield  {journal} {\bibinfo
  {journal} {Physics of Fluids B: Plasma Physics}\ }\textbf {\bibinfo {volume}
  {4}},\ \bibinfo {pages} {1152} (\bibinfo {year} {1992})}\BibitemShut
  {NoStop}%
\end{thebibliography}
